\documentstyle[psfig]{mn}

\newif\ifAMStwofonts
\AMStwofontstrue

\newcommand{\beq}{\begin{equation}}
\newcommand{\eeq}{\end{equation}}
\newcommand{\beqa}{\begin{eqnarray}}
\newcommand{\eeqa}{\end{eqnarray}}
\newcommand{\benu}{\begin{enumerate}}
\newcommand{\eenu}{\end{enumerate}}
\newcommand{\bite}{\begin{itemize}}
\newcommand{\eite}{\end{itemize}}
\newcommand{\bdes}{\begin{description}}
\newcommand{\edes}{\end{description}}
\newcommand{\refeq}[1]{equation (\protect\ref{#1})}
\newcommand{\reffig}[1]{Fig.\ \protect\ref{#1}}
\newcommand{\reftab}[1]{Table \protect\ref{#1}}
\newcommand{\refsec}[1]{Section \protect\ref{#1}}
\newcommand{\mv}{\mbox{$M_{V}$}}

\newcommand{\bv}{\mbox{$B-V$}}
\newcommand{\ub}{\mbox{$U-B$}}
\newcommand{\vk}{\mbox{$V-K$}}
\newcommand{\jk}{\mbox{$J-K$}}
\newcommand{\hk}{\mbox{$H-K$}}

\newcommand{\ubbv}{\mbox{(\ub)~vs.~(\bv)}}
\newcommand{\ubv}{\mbox{$UBV$}}

\newcommand{\feh}{\mbox{[Fe/H]}}

\newcommand{\Msolar}{\mbox{$M_{\odot}$}}
\newcommand{\Mpunto}{\mbox{$\dot{M}$}}
\newcommand{\Mcpunto}{\mbox{$\dot{M_{\rm c}}$}}
\newcommand{\sub}[1]{\mbox{$_{\rm #1}$}}
\newcommand{\Mi}{\mbox{$M\sub{i}$}}
\newcommand{\Mf}{\mbox{$M\sub{f}$}}
\newcommand{\Mto}{\mbox{$M\sub{TO}$}}
\newcommand{\Mhef}{\mbox{$M\sub{Hef}$}}
\newcommand{\Mup}{\mbox{$M\sub{up}$}}
\newcommand{\Mcore}{\mbox{$M\sub{c}$}}
\newcommand{\Menv}{\mbox{$M\sub{env}$}}
\newcommand{\McoreAGB}{\mbox{$M_{\rm c}^{\rm AGB}$}}
\newcommand{\McoreEAGB}{\mbox{$M_{\rm c}^{\rm EAGB}$}}
\newcommand{\McoreRGB}{\mbox{$M_{\rm c}^{\rm RGB}$}}
\newcommand{\McoreMS}{\mbox{$M_{\rm c}^{\rm MS}$}}
\newcommand{\Teff}{\mbox{$T\sub{eff}$}}
\newcommand{\diff}{\mbox{d}}

\title[\vk\ colours of stellar populations]
        {The evolution of the \vk\ colours 
        of single stellar populations }
\author[L. Girardi \& G.\ Bertelli]
       {L\'eo Girardi$^{1,2}$ and Gianpaolo Bertelli$^{3,4}$ \\
	$^1$Max-Planck-Institut f\"ur Astrophysik, 
	Karl-Schwarzschild-Str.\ 1, D-87540 Garching bei M\"unchen,
	Germany\\
	$^2$Alexander von Humboldt fellow \\
	$^3$National Council of Research (CNR-GNA) \\
	$^4$Department of Astronomy, Padova University, 
	Vicolo dell'Osservatorio 5, I-35122 Padova, Italy}
\date{Accepted 19?? ???.
      Received 1997 ???;
      in original form 1997 ???}

\pagerange{\pageref{firstpage}--\pageref{lastpage}}
\pubyear{199?}

\begin{document}

\maketitle

\label{firstpage}

\begin{abstract}
Models of evolutionary population synthesis of galaxies rely on the
properties of the so-called single stellar populations (SSP).  In this
paper, we discuss how the integrated near-infrared colours -- and
especially \vk\ -- of SSPs evolve with age and metallicity. Some of
the uncertainties associated to the properties of the underlying
stellar models are throughfully discussed.

Our models include all the relevant stellar evolutionary phases, with
particular attention being dedicated to the AGB, which plays a
fundamental role in the evolution of the near-infrared part of the
spectrum. First, we present the effects that
different formulations for the mass-loss rates produce on the final
remnant mass (i.e., on the initial--final mass relation), and hence on
the AGB-termination luminosity and the relative contribution of these
stars to the integrated light. The results for the evolution of the
\vk\ colour are very different depending on the choice of the
mass-loss prescription; the same happens also for the \bv\ colour in
the case of low-metallicity SSPs. Second, we describe the changes
occurring in the integrated colors at the onset of the AGB and RGB
stars. According to the classical formalism for the AGB evolution, the
onset of this evolutionary phase is marked by a color jump to the red,
whose amplitude is shown here to be highly dependent on the
metallicity and mass-loss rates adopted in the models. We then
consider the effect of the overluminosity with respect to the standard
core mass-luminosity relation, that occurs in the most massive AGB
stars. Different simplified formulations for this effect are tested in
the models; they cause a smoothing of the colour evolution in the age
range at which the AGB starts to develop, rather than a splitting of
the color jump into two separate events.  On the other hand, we find
that a temporary red phase takes place $\sim1.5\times10^8$~yr after
the RGB develops. Thanks to the transient nature of this feature, the
onset of the RGB is probably not able to cause marked features in the
spectral evolution of galaxies.

We then discuss the possible reasons for the transition of \vk\
colours (from $\sim1.5$ to 3) that takes place in LMC clusters of SWB
type IV (Persson et al.\ 1983).  A revision of the ages attributed to
the single clusters reveals that the transition may not be as fast as
originally suggested.  The comparison of the data with the models
indicates that it results mainly from the development of the AGB. A
gradual (or delayed) transition of the colours, as predicted by models
which include the overluminosity of the most massive AGB stars, seems
to describe better the data than the sudden colour-jump predicted by
classical models.

\end{abstract} 


\section{Introduction}

Models of evolutionary population synthesis are nowadays an important
tool for the study of nearby and distant galaxies. In them, the
spectral features observed in the integrated light of galaxies are
synthesized by summing the theoretical spectra of single stellar
populations (hereafter SSPs), as a way to get information into the
galactic history of stellar formation and chemical enrichment (e.g.\
Tinsley 1980; Bruzual \& Charlot 1993; Bressan, Chiosi \& Fagotto
1994b; Worthey 1994; Weiss, Peletier \& Matteucci 1995).

SSPs are defined as representative ensembles of stars of single age
and metallicity. They can be constructed by interpolation on extended
grids of stellar evolutionary tracks and spectra, and under the usual
assumption that the mass distribution of the stars is given by a known
initial mass function. The reliability of the SSP properties as a
function of their two fundamental parameters -- age and metallicity --
is secured by the previous testing of the stellar tracks and derived
isochrones against observations of stars in the solar neighbourhood
and in star clusters. Moreover, SSP models can be independently tested
by comparing their spectral properties with those observed in {\em
real} single-burst stellar populations. The best known examples of
these objects are star clusters, for which the internal spread of
stellar ages and metallicities is generally small enough to be
neglected.

The comparison between the properties of SSPs and star clusters has
been performed by several authors, mostly with relation to the
integrated broad-band colours (e.g.\ Chiosi, Bertelli \& Bressan 1988;
Arimoto \& Bica 1989; Battinelli \& Capuzzo-Dolcetta 1989; Barbaro \&
Olivi 1991; Girardi \& Bica 1993). This is a first step preliminary to
comparing spectral features. In a previous work (Girardi et al.\
1995), we showed that a particular set of models -- namely those from
the Padova group (see Bressan et al.\ 1994b; Bertelli et al.\ 1994) --
reproduces quite well two basic observational relations for the visual
colours, namely:

        1) {\em The linear relation between the $S$ parameter in the
\ubbv\ diagram and the logarithm of the age}, found in the star
clusters of the Magellanic Clouds by Elson \& Fall (1985). SSP models
with near-solar metallicity\footnote{In this paper we use the term
`near-solar metallicity' in opposition to `low-metallicity', which is
usually attributed to the globular clusters of the galactic halo. For
the sake of clarity, let us consider $Z=0.008$ (the metallicity of the
young populations in the Large Magellanic Cloud) as being the limit
between low and near-solar metallicity.} reproduce this linear
behaviour to a high extent, over most of the age interval that
comprehends young and intermediate-age star clusters.

        2) {\em The sequence of colours of galactic globular clusters}
by Racine (1973). The models for old SSPs (with $\sim15$ Gyr) are able
to reproduce both the slope and metallicity dependence of this
sequence in the \ubbv\ plane.  There are however, small systematic
shifts [of $\Delta(\bv)=0.05$ and $\Delta(\ub)=-0.03$ mag] between the
Padova models and the observations, which can be attributed mostly to
small inadequancies in the $\Teff$-colour transformations from Kurucz
(1992; see also Charlot, Worthey \& Bressan 1996).

This success in reproducing the observational relations is not an
exclusivity of the Padova models: similar behaviours for the \ubv\
colours can be found in several of the evolutionary population
synthesis models available in the literature. This is so because these
colours -- for both young and intermediate-age SSPs of near-solar
metallicity, and for old SSPs of low-metallicity, -- are mainly
determined by the stars in the well-understood evolutionary stage of
main sequence (MS).  These stars are known to evolve in a regular way
in the HR diagram, as a function of both age and metallicity. The
stars in the stage of core He-burning (CHeB) have a secondary
importance in determining the visual colours of SSPs. Although several
uncertainties persist in the theoretical predictions for this
evolutionary phase -- about e.g.\ the role of different convective
phenomena as core overshooting, semiconvection, and breathing pulses
(see Chiosi, Bertelli \& Bressan 1992) --, its behaviour in the HR
diagram is also reasonably regular. Moreover, its contribution to the
integrated colours is generally overwhelmed by that of the main
sequence. Thus, the CHeB stars are not able to change much the colour
behaviour dictated by the main sequence stars.  These considerations
are not valid for the very young SSPs of near-solar metallicity, since
for high-mass stars the evolution of the CHeB phase in the HR diagram
can significantly change with age and metallicity, and as a function
of the selected scheme of internal mixing. Higher uncertainties in the
predicted visual colours result in this case (cf.\ Girardi \& Bica
1993).

Near-infrared colours behave quite differently from the regular trends
found in the evolution of visual colours.  They are practically
determined by the evolution of the stars of later evolutionary phases,
especially those in the red and asymptotic giant branches (RGB and AGB
respectively), and by the red supergiants in the case of the youngest
SSPs. This leads to a non-regular evolution of the colours with age
and metallicity, since the amount and position of these stars in the
HR diagram is known to vary as a function of both parameters -- and
most remarkably as a function of age.

Let us focus our analysis on those stellar populations which contain
RGB and AGB stars. Stellar models predict that, as we go from higher
to lower stellar initial masses, these evolutionary phases appear
rather suddenly when we reach the upper mass limits for the
development of degenerate carbon-oxygen and helium cores, \Mup\ and
\Mhef, respectively.  The presence of these well-defined transition
masses then translate into major changes on the HR diagram properties
of star clusters over relatively small time scales. Renzini \& Buzzoni
(1986) concluded that the development of both AGB and RGB should be
accompanied by sudden jumps of the SSPs to redder integrated colors,
events which they called `phase transitions'. The concept of phase
transitions became very popular because they seemed to furnish a
plausible explanation for the bimodal distribution of the integrated
\bv\ colors of LMC clusters (see van den Bergh 1981), and because they
could provide useful `evolutionary clocks' for age-dating galaxies at
large redshifts (cf.\ Renzini \& Buzzoni 1986). However, Chiosi et
al.\ (1988) first concluded, with the aid of complete models of SSPs,
that the developments of the RGB is not accompanied by a noticeable
jump in the \bv\ colors, and that the bimodality in the distribution
of colors of LMC clusters could be attributed to other causes. The
latter conclusion was corroborated by the recent analysis of Girardi
et al.\ (1995). On the other hand, the colour jump due to the
development of the AGB was shown to be negligible in visual colours,
and was predicted to occur at too young ages to be useful in the
age-dating of distant galaxies (Bressan et al.\ 1994b).

However, the interest on the phase transition events has been renewed
by the suggestion of Renzini (1992) that the colour jump due to the
AGB phase transition could be severely delayed due to the effect of
envelope burning that occurs in the most massive AGB stars (Bl\"ocker
\& Sch\"onberner 1991). In this case, the AGB phase transition could
again be considered as a candidate for being a useful evolutionary
clock.  Moreover, since the effects of the AGB and RGB development
should be expected to be stronger in the near-infrared colours, they
were suggested to be at the origin of the apparently rapid transition
in $V-K$ colours, from $\sim1.5$ to 3, which occurs in the LMC
clusters of SWB type IV (see Renzini 1992; Corsi et al.\ 1994).  In
fact, the origin of this colour change remains to be properly
explained in the context of SSP models.

In this paper, we readdress the expected behaviour of the integrated
colours for those stellar populations that do contain AGB stars, with
the aim of casting light on the above mentioned points. Although our
main target is the study of the evolution of near-infrared colours --
mainly \vk\ --, we will also present the results for the \bv\ colour
for the sake of comparison.  In \refsec{sec_theory} we present an
overview of the methods used to calculate integrated colours of SSPs,
briefly considering what is the expected behaviour of the near-IR
colours, with base in the fuel consumption theorem by Renzini \&
Buzzoni (1986). Then we present the grid of stellar evolutionary
tracks which we use in order to calculate theoretical isochrones. They
include all evolutionary phases from the main sequence to the end of
the early (E-) AGB. The thermally pulsing (TP-) AGB is included by
means of a synthetic model. Some of the uncertainties, related to
the mass-loss rates and to the presence of `overluminous' AGB stars
with envelope burning, are thoroughfully discussed.  In
\refsec{sec_colours} we present the colour evolution based on the
isochrone method, discuss how it depends on several factors, and
revise the features that arise in the colours when the AGB and RGB
develop.  \refsec{sec_lmc} is dedicated to the study of the
near-infrared colours ($V-K$, $J-K$, and $H-K$) observed in Magellanic
Cloud star clusters. The results are briefly commented in
\refsec{sec_conclu}.

\section{Theoretical background and stellar models}
\label{sec_theory}

\subsection{The isochrone method and the fuel consumption theorem}
\label{sec_fct}

The simplest representation of single stellar populations is provided
by isochrones of given age and metallicity. These are produced from
stellar evolutionary tracks by means of a simple transformation from
the time evolution of stellar properties for several initial masses
[$f(t)_{M_{\rm i}}$], to the mass sequence of those properties for
several ages [$f(M_{\rm i})_t$].  The reliability of this kind of
transformation is secured by a fit mass grid of stellar tracks and a
suitable interpolation algorithm.

Once we have a set of isochrones, integrated colours of ideal SSPs can
be obtained by simply summing the contribution of different stars,
corresponding to different stellar masses, to the total light in a
given pass-band $\lambda$ (see e.g.\ Alongi \& Chiosi 1989; Girardi \&
Bica 1993; Charlot \& Bruzual 1991). For a given SSPs of age $t$ we
have
        \beq
L_{\lambda}^{\rm SSP}(t) = \int_{0}^{\infty} \phi_M
        L_{\lambda M}(t) \diff M
        \label{eq_isochrone}
        \eeq
where $L_{\lambda M}(t)$ corresponds to the luminosity along the
isochrone, and $\phi_M$ is the initial mass function.  The IMF is usually normalized so that the
total SSP mass is equal to a known quantity $M_{\rm T}$, i.e.\
        \[
        \int_{0}^{\infty} M\,\phi_M\,\diff M = M_{\rm T} \,.
        \]
Integrated magnitudes and colours follow straightforwardly from the
quantities $L_{\lambda}^{\rm SSP}(t)$.

This method of SSP construction is nowadays known as the `isochrone
method' (Charlot \& Bruzual 1991). Other approachs are also
possible, as discussed by e.g.\  Rocca-Volmerange
et al.\ (1996) and Lan\c con \& Rocca-Volmerange (1996). 
%One of them 
%is the `isomass method' (cf.\ Rocca-Volmerange
%et al.\ 1996). 
A useful approximation  
derives from the observation that in a SSP the
post-main sequence stars occupy a very limited interval of initial
masses.  Thus, all post-MS evolutionary phases can be described by a
single stellar track, of initial mass similar to the SSP turn-off mass
\Mto; and all MS stars can be represented by a single ZAMS line of
variable mass. With these approximations, \refeq{eq_isochrone} becomes
        \beqa
L_{\lambda}^{\rm SSP}(t) & = & \int_{0}^{M_{\rm TO}(t)} \phi_M 
        L_{\lambda M}(0) \diff M \nonumber\\
        & & + \, b(t) \,
        \int_{t_{\rm H}}^{\infty} L_{\lambda M_{\rm TO}}(\tau) \diff\tau \;,
        \label{eq_isomass}
        \eeqa
where the second integral correspond to the integration over the post-MS 
part (i.e.\ with $t>t_{\rm H}$) of the evolutionary track of mass \Mto, and
        \beq
b(t) = \phi_{M_{\rm TO}} 
        \left| \frac{\diff t\sub{H}}{\diff \Mi} 
		\right|_{M_{\rm i}=M_{\rm TO}}^{-1}
	\label{eq_evrate}
        \eeq
is the evolutionary flux, i.e., the number of stars which leave the MS
by unit time (see Renzini \& Buzzoni 1986). In the latter equation,
$t\sub{H}(\Mi)$ is the main sequence lifetime as a function of the
initial mass.

Equation \ref{eq_isomass}, when written for the bolometric luminosity
of a SSP, leads to the fuel consumption theorem as stated by Renzini
\& Buzzoni (1986): the second integral in the r.h.s.\ of
\refeq{eq_isomass} is replaced by the sum over a number $n$ of
evolutionary phases. Given the direct proportionality between the
quantity $L\Delta t$ for a single evolutionary phase $j$, and the
nuclear fuel $F_j$ consumed during it (the release of gravitational
energy and neutrinos being neglected), \refeq{eq_isomass} becomes
        \beq
L_{\rm bol}^{\rm SSP}(t) = L_{\rm bol}^{\rm MS}(t) +
        A \, b(t) \, \sum_{j=1,n} F_j(M_{\rm TO}) \,,
        \label{eq_fct}
        \eeq
where $L_{\rm bol}^{\rm MS}(t)$ is a well-behaved function of time
[see \refeq{eq_isomass}], and $A=9.75\times10^{10}\, L_{\odot}\, {\rm
yr}\, \Msolar^{-1}$ is a constant derived from the $Q$-value of
H-burning reactions.  The fuel consumption, in a first approximation,
is then given by
        \beq
F_j \simeq \Delta M\sub{H}_j + 0.1\,\Delta M\sub{He}_j \;,
        \label{eq_fuel}
        \eeq 
where $\Delta M\sub{H}_j$ and $\Delta M\sub{He}_j$ are the masses of H
and He, respectively, nuclearly burned during the $j$-th evolutionary
phase of the star of initial mass \Mto.

It can be noticed that the fuel consumption theorem involves a large
number of approximations, with respect to the original
\refeq{eq_isochrone}. However, it provides us with a powerful tool for
a qualitative understanding of the evolution of the integrated light
of SSPs, since it relates $L_{\rm bol}^{\rm SSP}(t)$ directly with the
quantities $F_j(M)$ derived from the stellar evolution theory.
Nowadays the use of \refeq{eq_isochrone} to calculate integrated
properties of SSPs is spread out due to the availability of several
different sets of isochrones in the literature, but we can still rely
on \refeq{eq_fct} in order to identify what causes the changes in the
integrated magnitudes and colours. This is, basically, the approach
adopted in this paper.

A concept useful to our analysis is that of the {\em total fuel
consumption}, here defined as the fuel consumption summed over all the
post-main sequence evolutionary phases:
        \beq
F_{\rm T} = \sum_{j=1,n} F_j .
        \label{eq_totalfuel}
        \eeq 
The post-AGB phases can be left out from this definition because their
fuel consumption is very small if compared to that from the preceding
evolutionary stages. For the stars that finish their nuclear evolution
as thermally pulsing AGB stars (i.e., for $\Mup>M\ga0.5$~\Msolar), the
intershell mass at the end of the AGB, $M\sub{H} - M\sub{He}$, takes
very small values if compared to \Mcore\ (typically, hundreths of
solar masses at its maximum values during quiescent phases of
H-burning, see e.g.\ Boothroyd \& Sackmann 1988), and then from
\refeq{eq_fuel} we have
        \beq 
F_{\rm T} \simeq 1.1 \McoreAGB - \McoreMS \,,
        \label{eq_agbfuel}
        \eeq 
where \McoreAGB\ and \McoreMS\ are the masses of the H-exhausted cores
at the end of the AGB and MS evolution, respectively. We recall that
\McoreMS\ is expected to be a well-behaving function of stellar
masses, similarly to $L_{\rm bol}^{\rm MS}$ which is a well-behaving
function of age.

The latter equation allows us to draw already some conclusions about
how the integrated near-infrared colours of SSPs vary with age.
Combined with \refeq{eq_fct}, \refeq{eq_agbfuel} tells us that, for
those SSPs that contain AGB stars (i.e.\ those with $\Mto\la\Mup$),
the evolution of the bolometric luminosity can be simply related to
the final core mass attained by them. 

We recall that the final core mass is also a quantity similar to the
final stellar (or remnant) mass, since at the end of the AGB evolution
only a very tiny envelope mass is left (of the order of thousanths of
solar masses, see Sch\"onberner 1981).

Therefore, we have an unique relation between the bolometric
luminosity of the SSPs, $L_{\rm bol}^{\rm SSP}(t)$, and the final
masses $\Mf\simeq\Mcore(t)$ of their AGB stars. If we have also a
single initial--final mass relation $\Mf(\Mi)$, and a simple,
monotonic relation between the initial mass and the SSP ages, then it
follows that {\em the evolution of the bolometric
luminosity of SSPs with $\Mto\la\Mup$ will closely reflect the
initial--final mass relation}.  The same holds, to some extent, for
the luminosities in the reddest photometric bands, since they sample
the region of the spectrum which is dominated by the coolest and more
luminous AGB stars, which are just those with the highest core
masses. Therefore, the number of these stars, and their relative
contribution to the near-infrared spectrum, are expected to vary also
in correspondence with $F_{\rm T}$. We can tentatively conclude that
the evolution of the near-infrared luminosity (e.g.\ the $K$ band),
and of colours envolving a visual and a near-infrared luminosity
(e.g.\ the \vk\ colour), should have some correspondence with the
behaviour of the initial--final mass relation.  We turn back to this
point in \refsec{sec_AGBpt}.

\subsection{Evolutionary tracks and AGB evolution}
\label{sec_tracks}

In this section we present the evolutionary tracks which we use in
order to compute the integrated colours of SSPs. We include an
overview of the properties which are relevant to the understanding of
the colour evolution, in the framework provided by the fuel
consumption theorem.

\subsubsection{From the MS to the end of the early-AGB}

We calculated two complete grids of stellar tracks of metallicities
$[Z=0.001, Y=0.23]$ and $[Z=0.008, Y=0.25]$, 
and with initial masses ranging from 0.15 to
7~\Msolar.  Initial stellar masses are spaced by 0.1~\Msolar\ over the
entire range of low-mass stars, and 1.0 \Msolar\ in the range from 3
to 7 \Msolar; this spacing is reduced to 0.05~\Msolar\ in the vicinity
of \Mhef, and to 0.5~\Msolar\ in the vicinity of \Mup. This mass
resolution is suitable to completely map the development of the tracks
in the HR diagram, as well as to follow the growth of the core mass at
the RGB-tip in the vicinity of \Mhef.  It allows us to study
the color evolution of SSPs even in some fine details.

The physical input of the models is as in the series of papers by
Bressan et al.\ (1994a) and Fagotto et al.\ (1994a,b), with some
updating in the equation of state, low-temperature opacities, and some
of the nuclear reaction rates (Girardi et al.\ 1996). For the sake of
brevity, we just recall here the main characteristics of the
models. They are calculated with recent OPAL (Iglesias \& Rogers 1996)
and low-temperature opacities (Alexander \& Ferguson 1994), and
moderate convective overshoot from stellar cores (Bressan et al.\ 1994a
and references therein). The equation of state includes both Coulomb
interactions in the center (cf.\ Girardi et al.\ 1996), and the H$_2$
molecule formation (cf.\ Mihalas et al.\ 1990 and references therein)
in the coolest stars.  The adoption of convective overshoot reflects
on the lower values of the limiting initial masses for building
degenerate cores after the exhaustion of central H and He: namely, we
find $\Mhef=1.9-2.0$~\Msolar, and $M\sub{up}=5$~\Msolar, whereas
classical models predict $\Mhef\sim2.2$~\Msolar\ and
$M\sub{up}=7-8$~\Msolar\ (e.g.\ Sweigart et al.\ 1990; Iben \& Renzini
1983).  Another point is that we avoid any artificial discontinuity in
the input parameters.  For instance, the overshooting parameter (see
Bressan et al.\ 1994a) is let to increase linearly from 0 up to its
maximum value, 0.5, in the mass range $1.0<(M/\Msolar)<1.5$. In this
way, the stellar lifetimes also behave as a smooth and monothonic
function of stellar mass.

In the case of low-mass stars, evolutionary tracks are interrupted at
the onset of the He-flash, and the corresponding HB track is continued
from a ZAHB model with the same core mass and composition of the last
RGB model.

Tables \ref{tab_z001} and \ref{tab_z008} present some of the
characteristics of the $Z=0.001$ and $Z=0.008$ models which may be
important to the considerations done in this paper.  Data on the
models with $0.15<(M/\Msolar)<0.6$ are not presented since these are
essentially non-evolving main sequence stars. Complete tables with the
evolutionary tracks and derived isochrones will be published
elsewhere, but are already available on request (e-mail:
\verb|leo@mpa-garching.mpg.de|).  
\begin{table}
\caption{Lifetimes and core masses for the $Z=0.001$ stellar tracks.}
\label{tab_z001}
\begin{tabular}{lcccc}
\noalign{\smallskip}\hline\noalign{\smallskip}
\Mi & $\tau_{\rm H}$ & \McoreRGB & $\tau_{\rm He}$ & \McoreEAGB \\
(\Msolar) & (yr) & (\Msolar) & (yr) & (\Msolar) \\
\noalign{\smallskip}\hline\noalign{\smallskip}
0.55        & $\ldots$ &      $\ldots$ & 1.266e8 & 0.4951 \\ 
0.60        & 40.87e9 &        0.4797 & 1.147e8 & 0.5030 \\
0.65        & $\ldots$ &      $\ldots$ & 1.136e8 & 0.5090 \\ 
0.70        & 22.51e9 &        0.4772 & 1.197e8 & 0.5139 \\
0.80        & 13.43e9 &        0.4762 & 1.181e8 & 0.5195 \\
0.90        & 8.477e9 &        0.4740 & 1.199e8 & 0.5247 \\
1.00        & 5.611e9 &        0.4721 & 1.072e8 & 0.5257 \\
%1.10        & 3.960e9 &        0.4711 & 1.079e8 & 0.5284 \\
%1.20        & 2.888e9 &        0.4698 & 1.158e8 & 0.5326 \\
%1.30        & 2.193e9 &        0.4691 & 1.062e8 & 0.5360 \\
%1.40        & 1.724e9 &        0.4676 & 1.126e8 & 0.5434 \\
%1.50        & 1.398e9 &        0.4660 & 1.127e8 & 0.5481 \\
1.10        & 3.903e9 &        0.4711 & 1.079e8 & 0.5284 \\
1.20        & 2.864e9 &        0.4698 & 1.158e8 & 0.5326 \\
1.30        & 2.256e9 &        0.4691 & 1.062e8 & 0.5360 \\
1.40        & 1.875e9 &        0.4676 & 1.126e8 & 0.5434 \\
1.50        & 1.594e9 &        0.4660 & 1.127e8 & 0.5481 \\
1.60        & 1.316e9 &        0.4359 & 1.247e8 & 0.5347 \\
1.65        & 1.207e9 &        0.4105 & 1.558e8 & 0.5310 \\
1.70        & 1.109e9 &        0.3431 & 2.341e8 & 0.5117 \\
1.75        & 1.021e9 &        0.3336 & 2.314e8 & 0.5281 \\
1.80        & 9.451e8 &        0.3289 & 2.222e8 & 0.5287 \\
1.90        & 8.177e8 &        0.3262 & 1.943e8 & 0.5472 \\
2.00        & 7.148e8 &        0.3300 & 1.714e8 & 0.5746 \\
2.20        & 5.615e8 &        0.3471 & 1.240e8 & 0.6186 \\
2.50        & 4.124e8 &        0.3835 & 0.767e8 & 0.6850 \\
3.00        & 2.739e8 &        0.4567 & 0.427e8 & 0.8132 \\
4.00        & 1.499e8 &        0.6254 & 0.183e8 & 0.8980 \\
4.50        & 1.174e8 &        0.7172 & 0.130e8 & 0.9345 \\ 
5.00        & 9.674e7 &        0.8206 & 0.976e7 & 0.9892 \\ 
5.50        & 7.948e7 &        0.9297 & 0.639e7 & 1.5053 \\ 
6.00        & 6.816e7 &        1.0419 & 0.584e7 & 1.7387 \\ 
7.00        & 5.092e7 &        1.2930 & 0.341e7 & 2.0302 \\ 
8.00        & 3.996e7 &        1.5741 & 0.249e7 & 2.4129 \\ 
\noalign{\smallskip}\hline\noalign{\smallskip}
\end{tabular}

\medskip \Mi\ is the initial mass; $\tau_{\rm H}$ and $\tau_{\rm He}$ are the 
lifetimes for central H and He-burning, respectively (`e' denotes a power of 
10); \McoreRGB\ and \McoreEAGB\ are the masses of the H-exhausted cores at the 
end of the RGB and E-AGB stages, respectively.\\ 

\end{table}

\begin{table}
\caption{The same as in \reftab{tab_z001}, but for $Z=0.008$.}
\label{tab_z008}
\begin{tabular}{lcccc}
\noalign{\smallskip}\hline\noalign{\smallskip}
\Mi & $\tau_{\rm H}$ & \McoreRGB & $\tau_{\rm He}$ & \McoreEAGB \\
(\Msolar) & (yr) & (\Msolar) & (yr) & (\Msolar) \\
\noalign{\smallskip}\hline\noalign{\smallskip}
0.55        & $\ldots$ &      $\ldots$ & 0.984e8 & 0.5103 \\ 
0.60        & 57.43e9 &        0.4518 & 1.007e8 & 0.5120 \\ 
0.70        & 32.85e9 &        0.4584 & 0.981e8 & 0.5187 \\
0.80        & 18.94e9 &        0.4707 & 0.965e8 & 0.5220 \\
0.90        & 11.88e9 &        0.4701 & 0.961e8 & 0.5245 \\
1.00        & 7.691e9 &        0.4693 & 0.941e8 & 0.5254 \\
1.10        & 5.064e9 &        0.4684 & 0.951e8 & 0.5280 \\
1.20        & 3.666e9 &        0.4684 & 0.947e8 & 0.5288 \\
1.30        & 2.945e9 &        0.4679 & 0.948e8 & 0.5311 \\
1.40        & 2.450e9 &        0.4676 & 0.938e8 & 0.5323 \\
1.50        & 2.077e9 &        0.4623 & 0.967e8 & 0.5312 \\
1.60        & 1.706e9 &        0.4529 & 1.020e8 & 0.5262 \\
1.70        & 1.426e9 &        0.4399 & 1.121e8 & 0.5234 \\
1.80        & 1.217e9 &        0.4155 & 1.330e8 & 0.5109 \\
1.85        & 1.127e9 &        0.4056 & 1.447e8 & 0.5094 \\
1.90        & 1.043e9 &        0.3243 & 2.910e8 & 0.4962 \\
2.00        & 9.097e8 &        0.3249 & 2.538e8 & 0.5001 \\
2.20        & 7.038e8 &        0.3284 & 2.110e8 & 0.5094 \\
2.50        & 5.088e8 &        0.3511 & 1.465e8 & 0.5407 \\
3.00        & 3.270e8 &        0.4102 & 0.690e8 & 0.6204 \\
3.50        & 2.289e8 &        0.4854 & 0.353e8 & 0.7574 \\
4.00        & 1.699e8 &        0.5668 & 0.203e8 & 0.8716 \\
4.50        & 1.317e8 &        0.6628 & 0.130e8 & 0.9177 \\ 
5.00        & 1.051e8 &        0.7668 & 0.905e7 & 1.1379 \\ 
5.20        & 9.656e7 &        0.8143 & 0.792e7 & 1.2125 \\       
5.50        & 8.596e7 &        0.8790 & 0.661e7 & 1.3275 \\ 
6.00        & 7.189e7 &        1.0052 & 0.505e7 & 1.5772 \\ 
7.00        & 5.264e7 &        1.2659 & 0.329e7 & 1.9626 \\ 
%8.00        & 3.996e7 &        1.5741 & 0.175e7 & 2.4129 \\ 
\noalign{\smallskip}\hline\noalign{\smallskip}
\end{tabular}

\end{table}

\subsubsection{The TP-AGB evolution}
\label{sec_tpagb}

We follow the TP-AGB evolution in an synthetic way (Iben \& Truran
1978), starting from the last AGB model calculated along the
evolutionary tracks. The total mass $M$, core mass \Mcore, effective
temperature \Teff, and luminosity $L$ of the models are let to evolve
according to some analytical formulas which inter-relate these
quantities and their time derivatives.  The formulas are basically a
subset of those collected by Groenewegen \& de Jong (1993) and Marigo
et al.\ (1996a). The end of the TP-AGB phase is determined either by
the total ejection of the envelope by stellar winds, or by the 
core mass growing up to the limit for explosive carbon ignition,
1.4 \Msolar\ (see however \refsec{sec_over}).

Without entering in the details about the integration procedure,
we list here the complete set of ingredients of our TP-AGB models:

\paragraph*{The core mass--luminosity relation (CMLR)} is from
Groenewegen \& de Jong (1993; and references therein):
        \beqa
L = \left\{ \begin{array}{ll}
        2.38\times10^5\, \mu\, Z\sub{CNO}^{0.04}\, 
                (\Mcore^2 - \\ \mbox{\hspace{0.5cm}} 
		0.0305\,\Mcore-0.1802)\:, 
                \mbox{\hspace{0.5cm}} \Mcore<0.7~\Msolar \\
        1.226\times10^5\, \mu^2\, (\Mcore-0.46)\, M^{0.19}\:, \\          
                \mbox{\hspace{0.5cm}} \Mcore>0.95~\Msolar
        \end{array} \right. 
        \label{eq_cmlr}
        \eeqa
where $\mu=4/(5X+3-Z)$ is the mean molecular weight, and $Z\sub{CNO}$
is the mass fraction of CNO elements in the stellar envelope, in our
case approximated by $0.71\,Z$. In the $0.95\le(\Mcore/\Msolar)\le0.7$
interval, $L$ is interpolated linearly between the above relations.

For the first thermal pulses, when the star has a luminosity below
that given by the CMLR, we use the same relation derived by Marigo et
al.\ (1996a) from the Vassiliadis \& Wood's (1993) models.  From the
last model of the evolutionary track, the luminosity and core mass of
are extrapolated according to the relation
        \beq
\frac{\diff L}{\diff\Mcore} = 60761\, \exp(M/2)
        \eeq
until $L$ equals the value given by the CMLR.

The breakdown of the CMLR, occurring in the AGB stars of higher masses
(Bl\"ocker \& Sch\"onberner 1991), will be throughly discussed in the
next section. This effect does not occur in stars of mass lower than
about 3~\Msolar, for which the above relations are valid.

\paragraph*{The evolutionary rate} \Mcpunto\ gives us the rate of 
core growth. According to Groenewegen \& de Jong (1993; and references
therein):
        \beq
\Mcpunto = 9.555\times10^{-12}\, \frac{L\sub{H}}{X}
        \eeq
where $L\sub{H}$ is the luminosity of the H-burning shell during the
quiescent phases of interpulse evolution, related to the total
luminosity by
        \beq
L = L\sub{H} + 2000(M/0.7)^{0.4} \exp[3.45(\Mcore-0.96)].
        \eeq

\paragraph*{The effective temperature} \Teff\ can also be given as a
function of basic stellar parameters. Given the simplicity of our
synthetic TP-AGB model, we need only the derivative of $\Teff(L)$,
because an initial value of the effective temperature is given by the
last evolutionary model we have in the complete stellar track. We
choose a $\Teff(L)$ relation with the same formal structure as that
used by Renzini \& Voli (1981), and differencing it with respect to
$L$ we obtain:
        \beq
\frac{\diff\log\Teff}{\diff\log L} = \gamma(Z) + 0.017 \,
        \frac{\diff\log M}{\diff\log L}. 
        \label{eq_teff}
        \eeq
The coefficient 0.017 in this equation is the one which better fit our
tracks for the early-AGB. Renzini \& Voli instead use a coefficient of
0.08.  The difference is probably due to the fact that the two
estimates are derived from stellar models calculated with very
different opacity tables and mixing length parameters.  The
$\gamma(Z)$ function is obtained directly from the slope on the HR
diagram of the complete evolutionary tracks at the stage of E-AGB,
evolved at constant mass and for different metallicities. An equation
like (\ref{eq_teff}) guarantees both the continuity of the AGB tracks
on the HR diagram, and a suitable dependence of the effective
temperature on the total mass of the star. On the other hand, it can
be an oversimplification of the real situation for TP-AGB stars of
higher luminosities, where the envelope structure (and hence \Teff)
can be strongly modified by the products of third dredge-up and by the
occurrence of hot bottom burning (e.g.\ Renzini \& Voli 1981; Marigo
et al.\ 1998).  We hope, however, that \refeq{eq_teff} can be an
useful first approximation for the kind of investigation we are
interested in.

\paragraph*{The mass-loss rate} \Mpunto\ constitutes one of the most
uncertain ingredients of the models of AGB stars. The classical
Reimers' (1975) formula,
        \beq
\Mpunto = 4\times10^{-13} \eta \frac{LR}{M} 
        \mbox{ ,\hspace{0.5cm} $\eta\sim1$}
        \label{eq_reimers}
        \eeq
(with $L$, $R$ and $M$ in solar units, here and throughout this
section) is known not to provide the high mass-loss rates observed in
real AGB stars, and required for the production of planetary nebulae
at the end of the AGB (Iben \& Renzini 1983). We consider here 3 other
formulations for the mass-loss rates, all of them sharing the
characteristic of developing a `superwind' phase (with
$\Mpunto\ga10^{-5}$ \Msolar/yr) at moderate luminosities. It follows a
brief description of them.

\benu

\item The Bowen \& Willson (1991) prescription is based on the results
of hydrodynamical models for the envelope of giant pulsating stars
(Bowen 1988).  It is characterized by an exponential increase of the
mass-loss rate with the luminosity, or equivalently, with the
pulsating period $P$. The Bowen \& Willson's results are not presented
in the form of analytical formulas for the mass-loss rate as a
function of relevant stellar parameters.  However, the following
formula is a fit to the results presented in their Fig.~2:
        \beqa
\log\Mpunto = \left\{ \begin{array}{ll}
        -4.925\log M + 7.65\log P - 26.08\:, \\
                \mbox{\hspace{0.5cm}} M\le1.2~\Msolar \\
        -24.996\log M + 7.08\log P + \\
                \mbox{\hspace{0.5cm}} 7.198\log M\log P - 24.489\:, \\  
                \mbox{\hspace{0.5cm}} M\ge1.2~\Msolar
        \end{array} \right. 
        \label{eq_bowen}
        \eeqa 
This equation gives values for \Mpunto\ very close to those originally
presented by Bowen \& Willson (the maximum differences are of
only $\Delta\log\Mpunto=-0.14$). The pulsation period $P$ (in days),
can be obtained as a function of basic stellar parameters by means of
Vassiliadis \& Wood's (1993) period--mass--radius relation:
        \beq 
\log P = -2.07 + 1.94\log R - 0.9\log M  \,. 
        \label{eq_periodo} 
        \eeq 

\item The Bl\"ocker's (1995) prescription was constructed with the aim of 
expressing the theoretical results of Bowen (1988) by means of a simple 
approximative formula:
        \beq
\Mpunto = 4.83\times10^{-9}\, L^{2.7}\, M^{-2.1} \Mpunto\sub{R},
        \label{eq_blocker}
        \eeq
where $\Mpunto\sub{R}$ is the Reimers' mass-loss rate
[\refeq{eq_reimers}] with $\eta=1$. In fact, comparing the results of
equations (\ref{eq_bowen}) and (\ref{eq_blocker}) we notice that the
latter gives values for \Mpunto\ that are typically much higher than
those originally presented in Bowen's work.

\item Vassiliadis \& Wood (1993) give a prescription for the mass-loss
rates which is mostly based on observational data for long-period
variables (Mira and OH/IR) stars. It relates \Mpunto\ with the
pulsation period $P$ (in days):
        \beq
\log\Mpunto = -11.4 + 0.0123\, P   \mbox{,\hspace{0.5cm} $P\la500$ days.}
        \label{eq_vw}
        \eeq
For periods longer than $\sim500$ days, the mass-loss rate is given by
the theory of radiation-driven winds, or
        \beq
\Mpunto = 6.07\times10^{-3} \frac{L}{c\, v\sub{exp}},
        \label{eq_rdw}
        \eeq
where $c$ is the velocity of light, and $v\sub{exp}$ is the wind
expansion velocity, $v\sub{exp}=-13.5+0.056\,P$, both in km$\,{\rm
s}^{-1}$.  In our calculations the adopted mass-loss rate is the
highest one between those given by (\ref{eq_vw}) and (\ref{eq_rdw}).

\eenu

\paragraph*{}

The above relations are enough to describe the time evolution of the
stellar parameters $L$, \Teff, \Mcore\ and $M$ along the TP-AGB. It is
worth recalling that this TP-AGB model is much simpler than, and
probably not as accurate as, those presented in some recent papers in
the literature (e.g.\ Groenewegen \& de Jong 1993; Marigo et al.\
1996a).  However, our theoretical scheme is suitable to the main scope
of this paper, which is to give an appropriate description of how the
integrated colours of SSPs evolve with time. We notice, for example,
that the results for the initial--final mass relation obtained by us
adopting Vassiliadis \& Wood's (1993) mass-loss rates, and in the case
of $Z=0.008$, are very similar to those derived by Marigo et al.\
(1996a). This despite the fact that the latter authors use a much more
detailed TP-AGB model, which accounts for various effects such as the
changes in the chemical composition at the stellar surface, and makes
use of a complete envelope model in order to obtain the effective
temperatures.  This agreement is encouraging, since the final core
mass is a key-quantity for our colour evolution models [see
\refeq{eq_totalfuel}].

\begin{figure*}
%\vspace{8.3cm}
\begin{minipage}{8.3cm}
\psfig{file=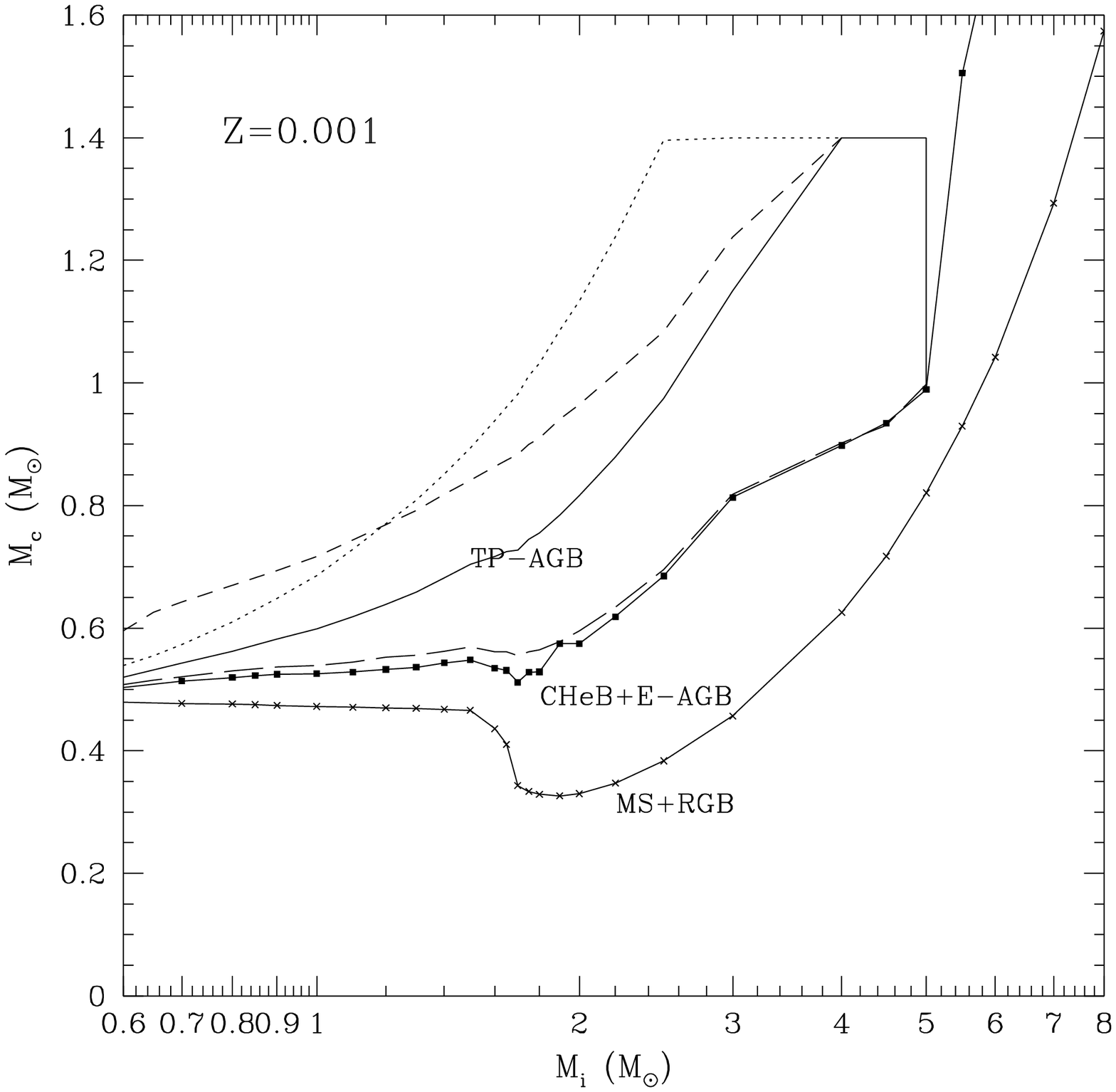,width=8.3cm}
\end{minipage}
\hfill
\begin{minipage}{8.3cm}
\psfig{file=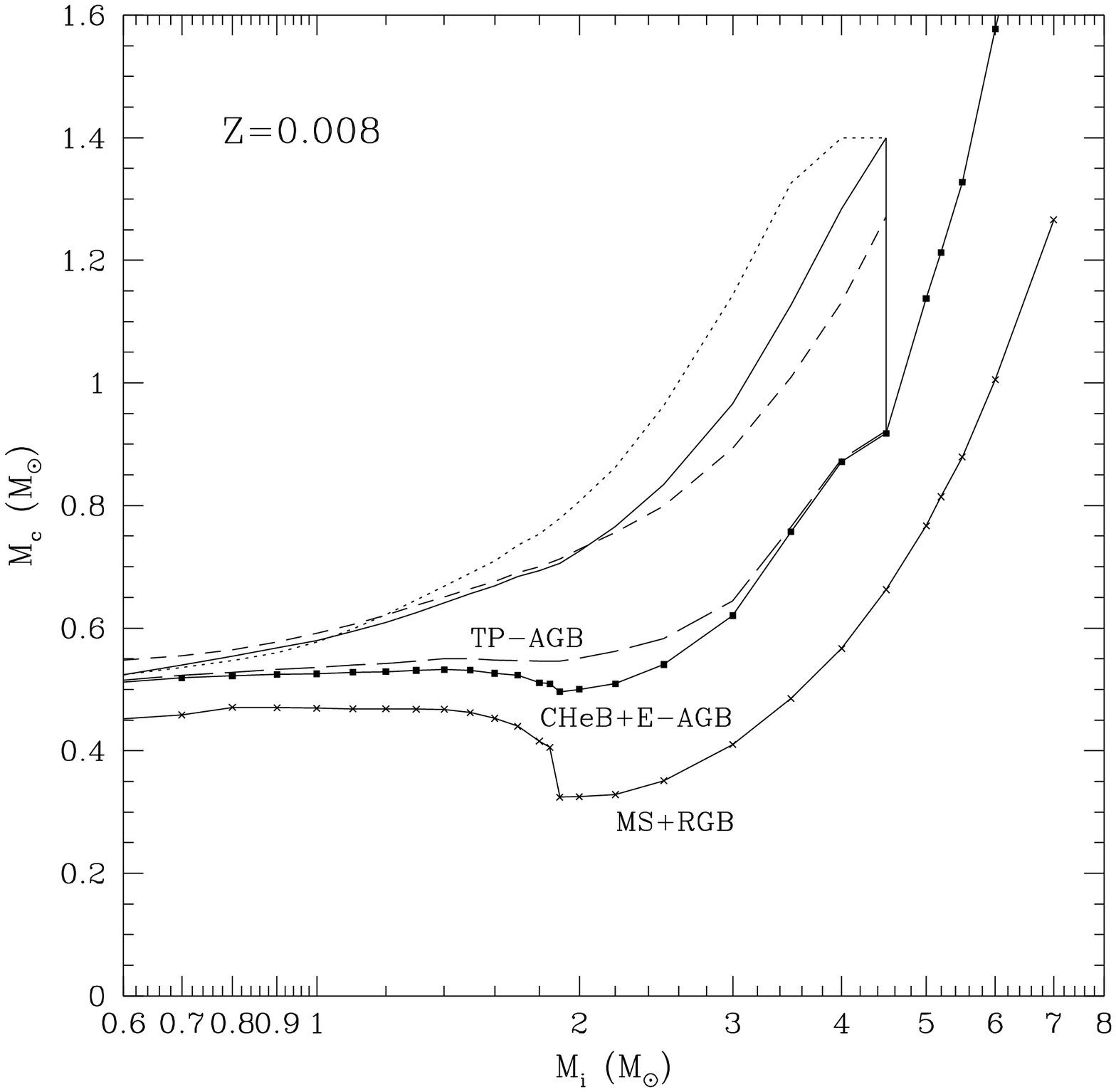,width=8.3cm}
\end{minipage}
        \caption{  
The He-core mass as a function of initial mass, at both the start of
CHeB (RGB-tip for low-mass stars, or quiet He-ignition for
intermediate-mass stars), and at the TP-AGB-end, for initial
metallicities $Z=0.001$ (left panel) and $Z=0.008$ (right panel). The
lines are as follows: the lower continuous line with crosses is the
He-core mass at the beginning of the CHeB phase, \McoreRGB; the
continuous line with black dots immediately above is the core mass at
the start of the thermally pulsing regime on the AGB, \McoreEAGB, or
quiet C-ignition for high-mass stars; then follow the core mass at the
end of AGB evolution (complete ejection of the envelope or supernova
explosion) for 4 different prescriptions of mass-loss: Reimers (1975)
with $\eta=1.0$ (continuous line), Bowen \& Willson (1991; dotted
line), Vassiliadis \& Wood (1993; short-dashed line) and Bl\"ocker
(1995; long-dashed line).
        } 
\label{fig_mcore} 
\end{figure*}

\subsubsection{The core mass at several evolutionary stages}
\label{sec_core_mass}

The final core mass built up at the end of the AGB, \McoreAGB, is
highly dependent on the assumed mass-loss
rates. Figure~\ref{fig_mcore} shows the values of \McoreAGB\ for the 4
different formulations for the mass-loss presented in the above
section.

In \reffig{fig_mcore} it is clear that the Vassiliadis \& Wood (1993)
and Bowen \& Willson's (1991) mass-loss formulas, contrarily to the
Reimer's (1975) one, give origin to a strong metallicity dependence of
the final core mass attained by intermediate-mass stars. In this way,
stars of lower metallicity finally develop higher core masses,
and can easily attain the
critical core mass of 1.4~\Msolar. This would result in supernova
explosions of type I$\,1/2$ (Iben \& Renzini 1983), with important
implications for the theory of chemical evolution of galaxies (see
Matteucci \& Tornamb\`e 1985). The Bl\"ocker's (1995) formulation,
instead, gives origin to very high values of mass-loss rates as soon
as it is activated in the synthetic TP-AGB calculation.  For this
reason, in \reffig{fig_mcore} the final core masses developed by
models with this choice of mass-loss, are very close to the core
masses at the start of thermal pulses. It stresses our conclusion that
Blocker's formulation for the mass-loss rates is indeed very different
from the original ones obtained by Bowen (1988) and Bowen \& Willson
(1991).

We note on passing that models which use Vassiliadis \& Wood's (1993)
mass-loss rates were already shown to reproduce the maximum luminosity
of AGB stars as a function of age in Magellanic Cloud clusters (cf.\
Marigo, Girardi \& Chiosi 1996b), for ages $8.8\la\log(t/\rm
yr)\la9.6$. This provides also a useful constraint to mass-loss rates,
since $L$ and \Mcore\ are straightly linked by the CMLR. For this
reason, and because it is based on observational data, we believe that
Vassiliadis \& Wood's formulation for \Mpunto\ is probably the most
realistic between those tested in this work. A conclusive test on the
mass-loss rates is provided by a comparison of models with solar
metallicity and the initial--final mass relation observed in the solar
vicinity (Herwig 1996). In fact, tests made
with $Z=0.02$ models revealed that TP-AGB stars with Reimers' (1975)
mass-loss rates end their evolution with too large core masses and
luminosities, while those with Bl\"ocker's (1995) mass-loss rates end
with too small core masses.

An important aspect to be noticed in \reffig{fig_mcore} and
\reftab{tab_z001} is that
        \bdes
        \item -- 
the core mass at the beginning of the He-burning (corresponding to the
RGB-tip for low-mass stars and to quiet He-ignition for
intermediate-and high-mass stars), \McoreRGB, is a non-smooth function
of initial mass. It presents the well-known plateau of
$\Mcore\simeq0.45$~\Msolar\ for $\Mi<\Mhef$, a rapid jump from
$\sim0.45$ to $\sim0.32$ \Msolar\ in the vicinity of $\Mi\simeq\Mhef$,
and increases monotonically with the stellar mass for $\Mi>\Mhef$;
        \item --
on the contrary, irrespective of the adopted mass-loss formulation,
{\em the final core mass attained by AGB stars, \McoreAGB, is a smooth
and monotonically increasing function of the initial stellar mass} for
the entire range of $\Mi<\Mup$.
        \edes
It means that the He-burning evolution has the property of almost
completely erasing the memory about the core mass at the earlier
evolutionary stages. Most of this leveling-off effect takes place
during the HB and E-AGB evolution, as can be noticed by looking at the
lines in \reffig{fig_mcore}.  The residual discontinuities in the
function $\Mcore(\Mi)$ are smoothed out during the TP-AGB, as a
natural result of
        \benu
        \item the existence of a core mass--luminosity relation during
most of the TP-AGB evolution,
        \item of the fact that all mass-loss formulas are simply
expressed as a regular function of fundamental stellar parameters as
the effective temperature and luminosity, and
        \item of this mass-loss being a very steep function of the 
luminosity. 
        \eenu
Considering the straight relation between the final core mass and the
total fuel consumption of a star [cf.\ \refeq{eq_agbfuel}], it implies
that the total fuel consumption is also a smooth and monotonic
function of the initial mass, even in the mass range in which the
development of the RGB starts.

\subsubsection{The effect of AGB overluminosity}
\label{sec_over}
 
It is nowadays clear that the core mass-luminosity relation does not
describe the luminosity evolution of all AGB stars. Tuchman, Glasner
\& Barkat (1983) clarified that the validity of the CMLR depends on
the existence of a radiative region between the H-burning shell and
the convective envelope.  For the AGB stars of higher masses,
significant burning through the CNO cycle can occur at the bottom of
the convective envelope, thus breaking down the validity of the CMLR
for these stars, and allowing their luminosities to grow apparently
unconstrained (Bl\"ocker \& Sch\"onberner 1991; Boothroyd \& Sackmann
1992). We will refer to this behaviour as the `AGB overluminosity'.
Its effect on AGB evolution can be qualitatively understood as
follows:
        \benu
        \item At the beginning of the TP-AGB, the star starts nuclear
burning at the bottom of the envelope, and grows rapidly in luminosity
ignoring the limitation imposed by the CMLR;
        \item the high luminosities soon trigger high mass-loss rates;
        \item once a significant amount of stellar mass has been lost,
the conditions for envelope burning being active no longer exist, and
the star goes down in luminosity up to attaining the CMLR;
        \item the star evolves along the CMLR, just as if it were an
AGB star of lower initial mass. Conditions for reactivating the
envelope burning are probably not met again up to the end of its life.
        \eenu
This kind of behaviour can be found e.g.\ in the 5~\Msolar\ models
computed by Vassiliadis \& Wood (1993), in those of Boothroyd \&
Sackmann (1992), Bl\"ocker (1995), and more recently in Marigo et al.\
(1998). The net result, according to the simplified picture above
presented, is that AGB stars with high initial mass pass through a
temporary phase of high luminosity, but
soon after settle on an evolutionary behaviour which is in many
aspects similar to those of AGB stars born with lower initial masses.
Thus, the final core mass attained by these stars is {\em lower} than
that which would be attained if the CMLR were always valid for
them. The same holds for the total fuel consumption, according to
\refeq{eq_agbfuel}. Notice that a probable consequence of the
overluminosity effect is that of preventing the core mass of AGB stars
of growing up to the limit of 1.4 \Msolar, thus accounting for the
non-observation of supernovae of type I$\,1/2$, even at very low
metallicities.

The consideration of the overluminosity effect in synthetic
calculations of the TP-AGB evolution is not an easy task.  Complete
evolutionary calculations indicate that the efficience of envelope
burning, as well as the amount of overluminosity, depends critically
on a series of parameters as the envelope mass, core mass, metallicity
and mixing length parameter (Sackmann \& Boothroyd 1991; Boothroyd \&
Sackmann 1992; Boothroyd, Sackmann \& Ahern 1993).  At present, the
only attempt to provide analytical formulas feasible to describe the
luminosity increase of envelope burning stars, as a function of these
parameters, is that from Wagenhuber \& Groenewegen (1997).

In order to test the effect that the overluminosity effect can have on
integrated colours of SSPs, we would like to present here calculations
made under very simplified assumptions. We think that these
experiments are worth because, up to this moment, the overluminosity
effect was not considered in this kind of models. The simplest
assumptions we can work with are that the overluminosity takes place
in one of the following cases:
        \benu
        \item when the core mass is higher than 0.85 \Msolar;
	\label{item_bs}
        \item when the envelope mass is higher than 1.6 \Msolar.
	\label{item_vw}
        \eenu
Case \ref{item_bs} corresponds to the statement originally made by
Bl\"ocker \& Sch\"onberner (1991) and Boothroyd \& Sackmann (1992). It
dates from the very first papers which dealt with envelope burning
stars. It is now clear that this rule does not hold always: for
instance, the Vassiliadis \& Wood's (1993) models with an initial mass
of 5~\Msolar\ have $\Mcore\ga0.95$ \Msolar\ and still follow a CMLR at
the end of their evolution. These authors conclude that below certain
values of envelope mass the hot bottom burning does not occur. This
result is represented by the case \ref{item_vw} above.

\begin{figure*}
%\vspace{8.3cm}
\begin{minipage}{8.3cm}
\psfig{file=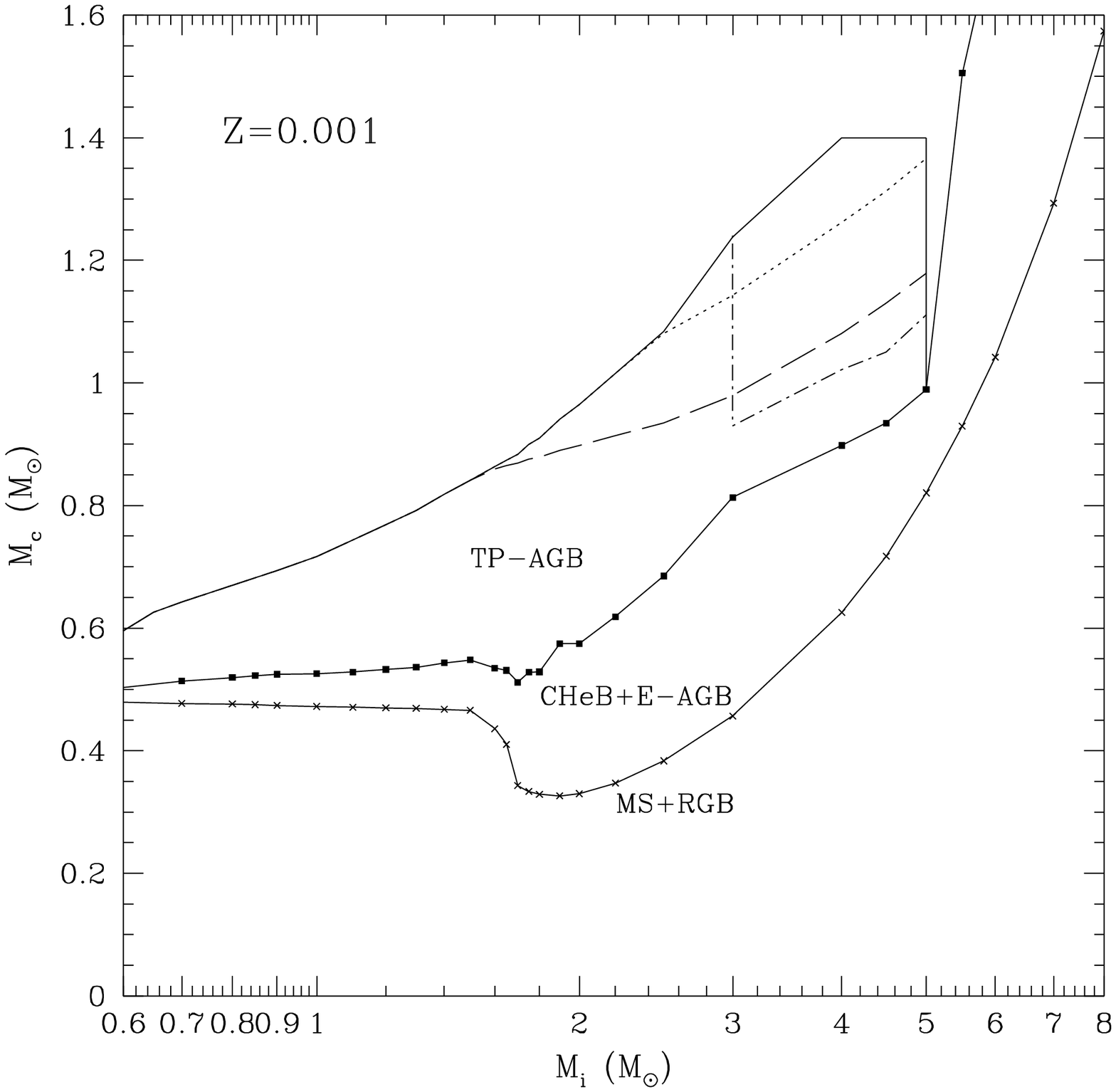,width=8.3cm}
\end{minipage}
\hfill
\begin{minipage}{8.3cm}
\psfig{file=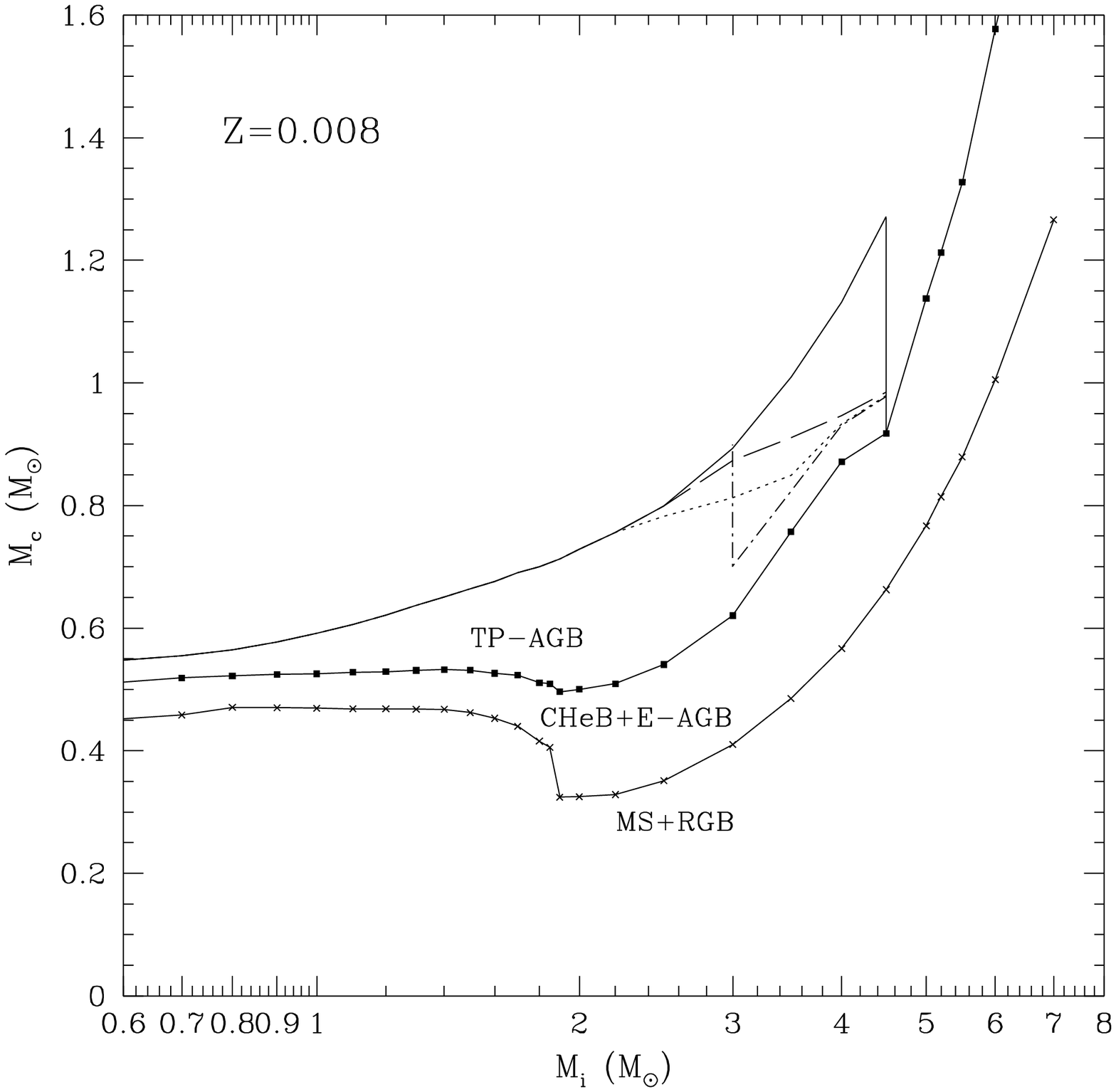,width=8.3cm}
\end{minipage}
        \caption{  
The same as \reffig{fig_mcore}, but limited to the case of Vassiliadis
\& Wood's (1993) mass-loss rates, and considering the overluminosity
effect on the most massive AGB stars according to 2 different
prescriptions: as if the envelope mass were the only parameter
determining the overluminosity [\refeq{eq_overa}; dotted line], and as
if the core mass also plays a role in the overluminosity effect
[\refeq{eq_overb}; long-dashed line].  For comparison, the curve
obtained with the assumption of the validity of the CMLR
(\reffig{fig_mcore}) is presented as a continuous line.  The
dotted-dashed line, instead, illustrates the predictions by Renzini
(1992), according to whom the overluminosity effect reduces
drastically the AGB lifetime for stars above a certain initial mass
value. At sufficiently low masses, all the different model present the
same final core masses.
        } 
\label{fig_mcoreover} 
\end{figure*} 

Let us consider that the envelope mass is the main determinant factor
of the overluminosity, and assume a value of 50 percent of
overluminosity for envelope masses in excess of 0.4 \Msolar, so we
have:
        \beqa
\frac{L}{L\sub{CMLR}} & = & 1 + 0.5 \frac{(M\sub{A}/\Msolar-1.6)}{0.4} 
        \:, \nonumber\\
        & & \mbox{for } M\sub{A}= {\rm max}(\Menv,1.6\,\Msolar) \,.
        \label{eq_overa}
        \eeqa
As an alternative formulation, we consider core masses in excess of 0.85 
\Msolar\ as a concurrent factor for the overluminosity: 
        \beqa
\frac{L}{L\sub{CMLR}} & = & 1 + 
        0.25 \frac{(M\sub{B}/\Msolar-0.85)}{0.1}
        \frac{(M\sub{A}/\Msolar-1.6)}{0.4} \:, \nonumber\\
        & & \mbox{for } M\sub{B}= {\rm max}(\Mcore,0.85\,\Msolar) \,.
        \label{eq_overb}
        \eeqa

Introducing these prescriptions in our equations for the stellar
luminosities, we generate synthetic AGB models which attain the core
masses illustrated in \reffig{fig_mcoreover}. We plot only the results
obtained for the Vassiliadis \& Wood (1993) mass-loss
formulation. It can be noticed that
these prescriptions for the overluminosity cause a flattening of the
initial--final mass relation for the most massive intermediate-mass
stars (with $\Mup\ga\Mi\ga2.5$~\Msolar\ for $Z=0.008$).  We recall
that some authors have suggested that the initial--final mass relation
observed in the solar vicinity presents some evidences of such a
flattening (see Herwig 1996; Jeffries 1997).  For the stars in the
vicinity of \Mup, the AGB lifetime is shortened by a factor of up to
5. For the stars of lower masses, the overluminosity effect causes a
progressive reduction of the AGB lifetime.  For $\Mi\la2.5$ \Msolar,
the CMLR results to be valid through the entire AGB evolution.

The amount of overluminosities predicted by equations (\ref{eq_overa})
and (\ref{eq_overb}) are quite high if confronted to those found by
several authors. However, we preferred to explore the most extreme
possible conditions, since the results that would be obtained with
more conservative assumptions, of course, would approach the results
obtained in the case of no overluminosity. The reader should keep in
mind the possibility of intermediate behaviours for the curves
depicted in \reffig{fig_mcoreover}, which would also reflect on the
results presented below.

\subsubsection{Theoretical isochrones}
\label{sec_isochrones}

Isochrones are constructed from the stellar tracks by means of a
simple interpolating algorithm. The smooth and regular behaviour of
the isochrones in the HR diagram is guaranteed by interpolating
between points of equivalent evolutionary status over the stellar
tracks, and by using $\log\Mi$ and $\log t$ as the variables of
interpolation. The procedure is similar to that used by Bertelli et
al.\ (1994).  The transformation from the theoretical to the observed
stellar quantities is also performed as in Bertelli et al.\ (1994), to
whom we refer for all details. Suffice it to say that they rely on the
Kurucz's (1992) library of stellar spectra, complemented at low
effective temperatures with observed spectral energy distributions
from several authors. Giant stars are attributed \vk\ colours
according to Bessell \& Brett (1988) and Terndrup et al.\ (1991), and
the \Teff\ scale from Lan\c con \& Rocca-Volmerange (1992).

\section{Evolution of integrated colours}
\label{sec_colours}

We now start discussing how the integrated colours evolve, in the
different cases previously described. We recall that our models contain
all the evolutionary phases relevant for the determination of the
optical and near-infrared colours of SSPs.

\begin{figure*}
\begin{minipage}{8.3cm}
\psfig{file=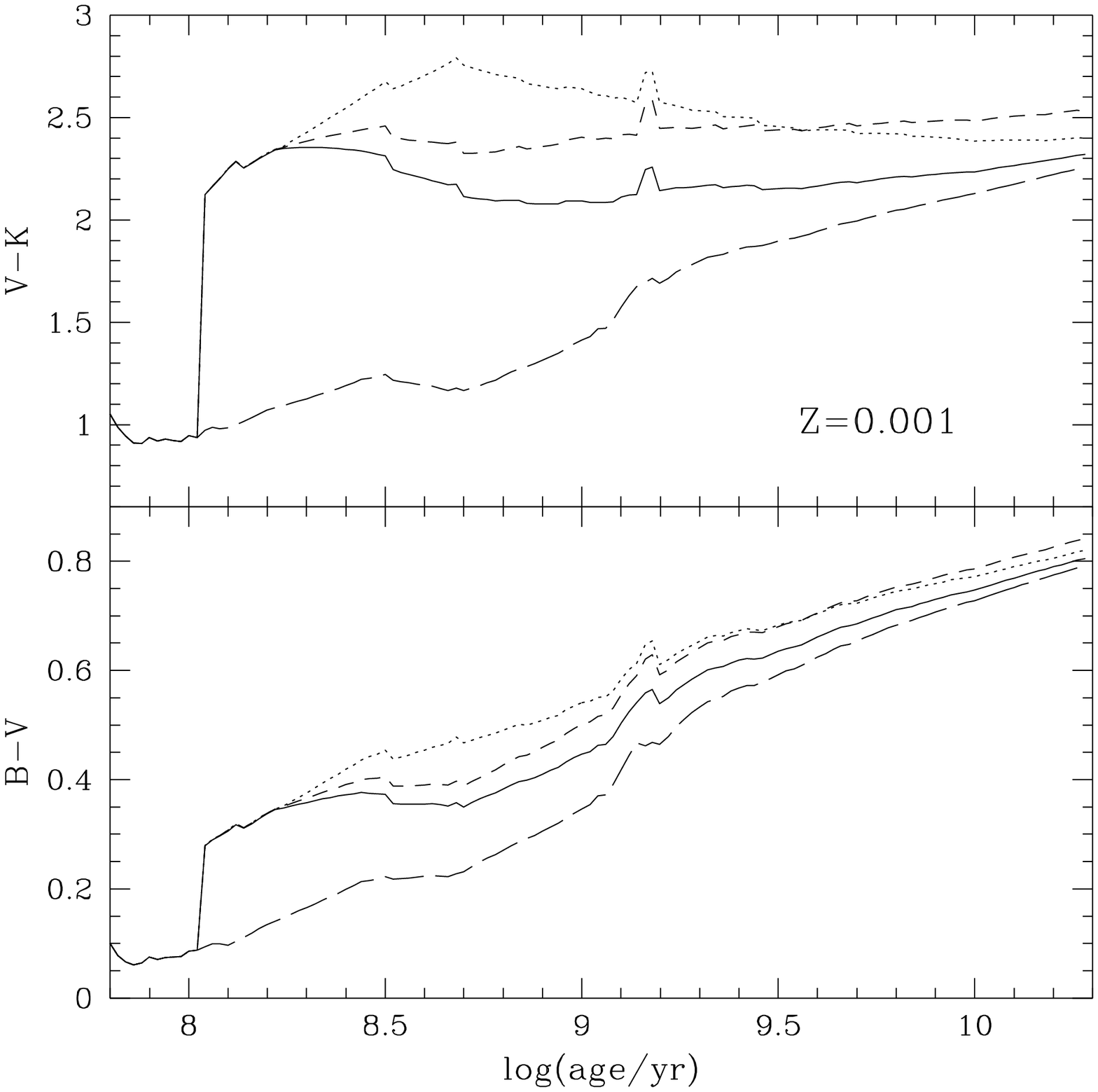,width=8.3cm}
\end{minipage}
\hfill
\begin{minipage}{8.3cm}
\psfig{file=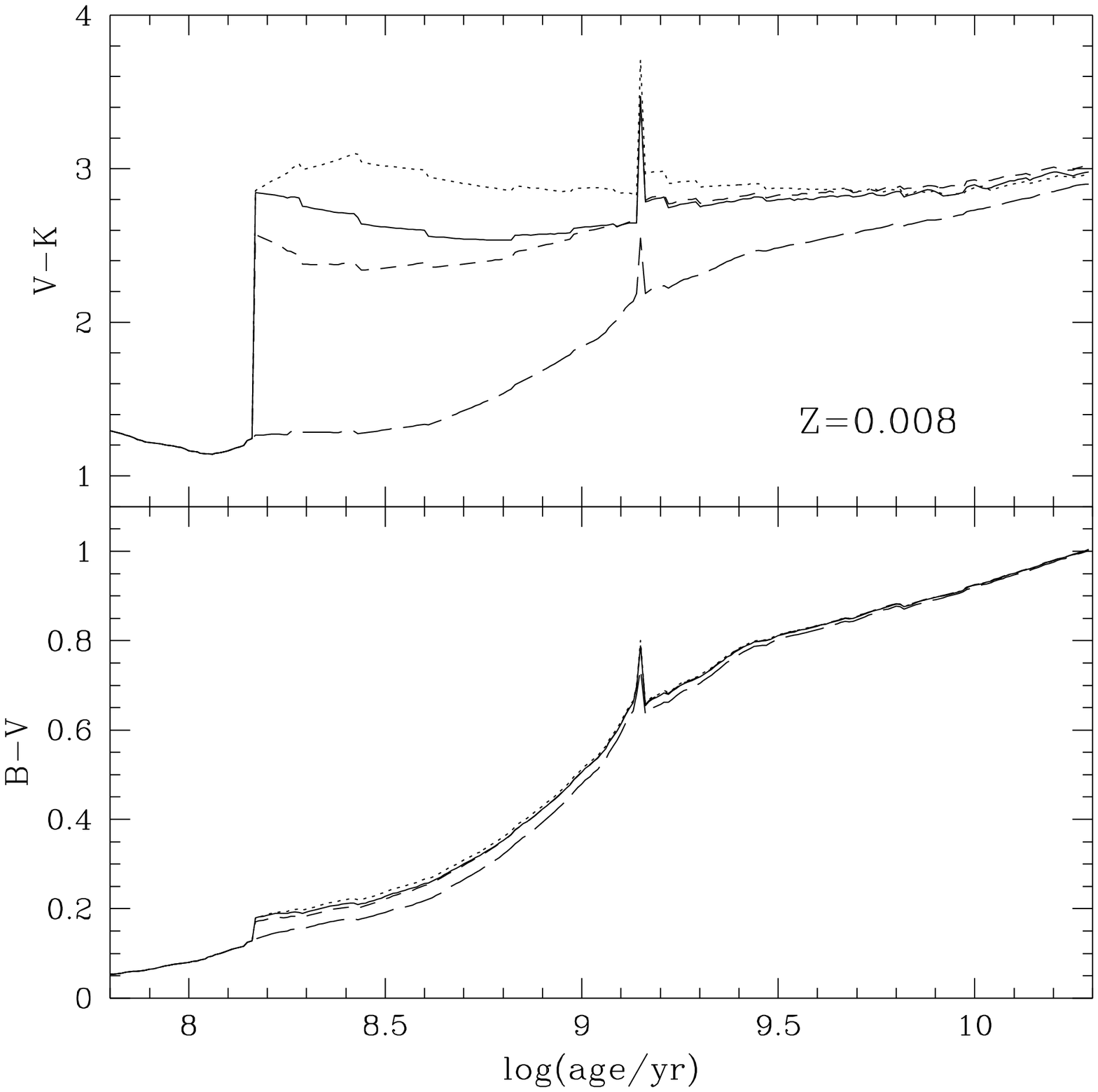,width=8.3cm}
\end{minipage}
        \caption{  
The time evolution of the \bv\ and \vk\ colours, for SSPs with
metallicities $Z=0.001$ and $Z=0.008$, and according to 4 different
choices for mass-loss rates along the TP-AGB: Reimers (1975) with
$\eta=1.0$ (continuous line), Bowen \& Willson (1991; dotted line),
Vassiliadis \& Wood (1993; short-dashed line) and Bl\"ocker (1995;
long-dashed-line). The age interval shown covers entirely (and
is slightly larger) that in which AGB stars are present in the SSPs.
The features in the colour evolution at ages slightly larger than 
$10^8$ and $10^9$~yr, are associated with the onset of AGB and RGB
stars, respectively (see text for details).
        }
\label{fig_colori}
\end{figure*} 

\begin{figure*}
\begin{minipage}{8.3cm}
\psfig{file=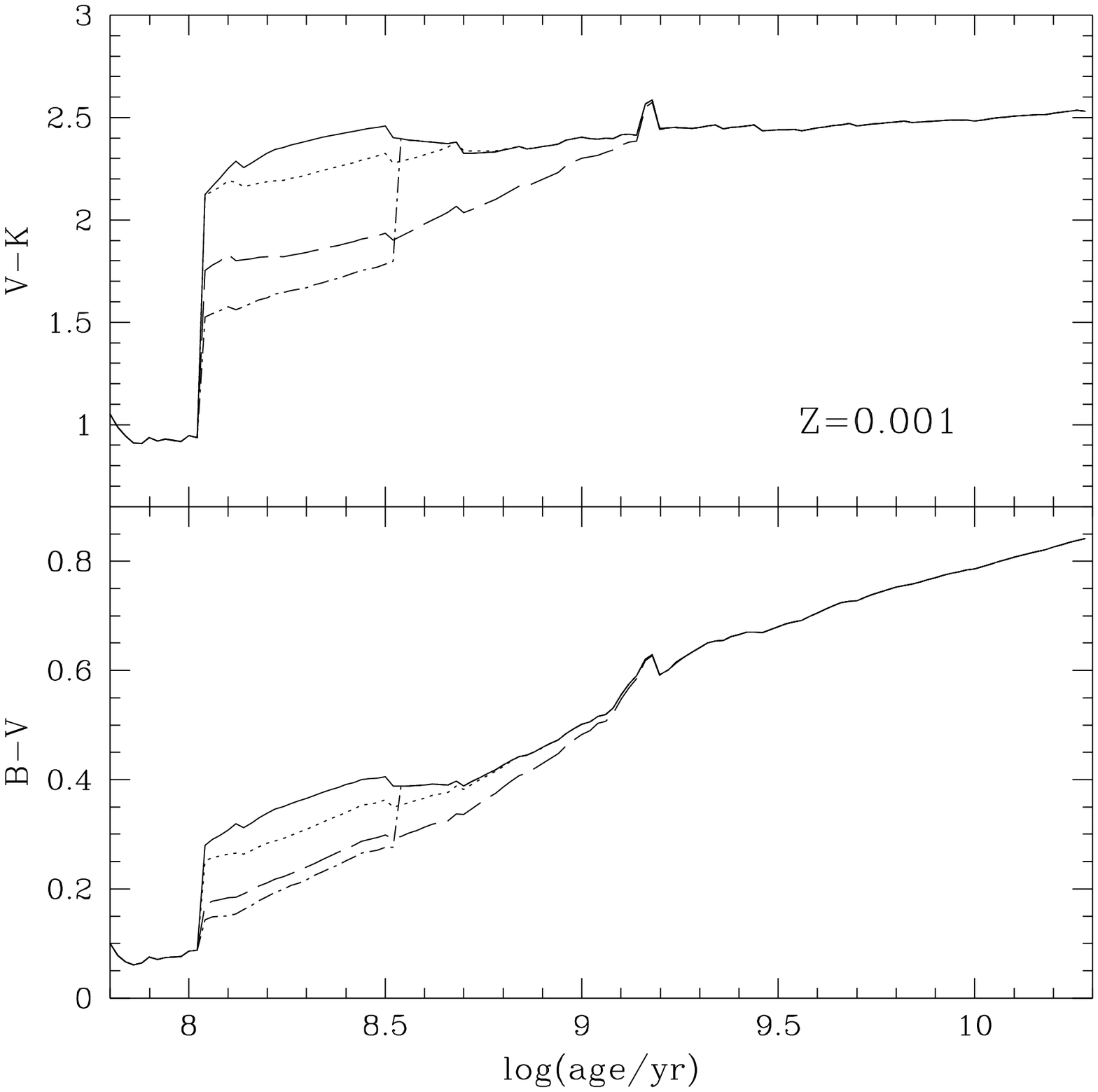,width=8.3cm}
\end{minipage}
\hfill
\begin{minipage}{8.3cm}
\psfig{file=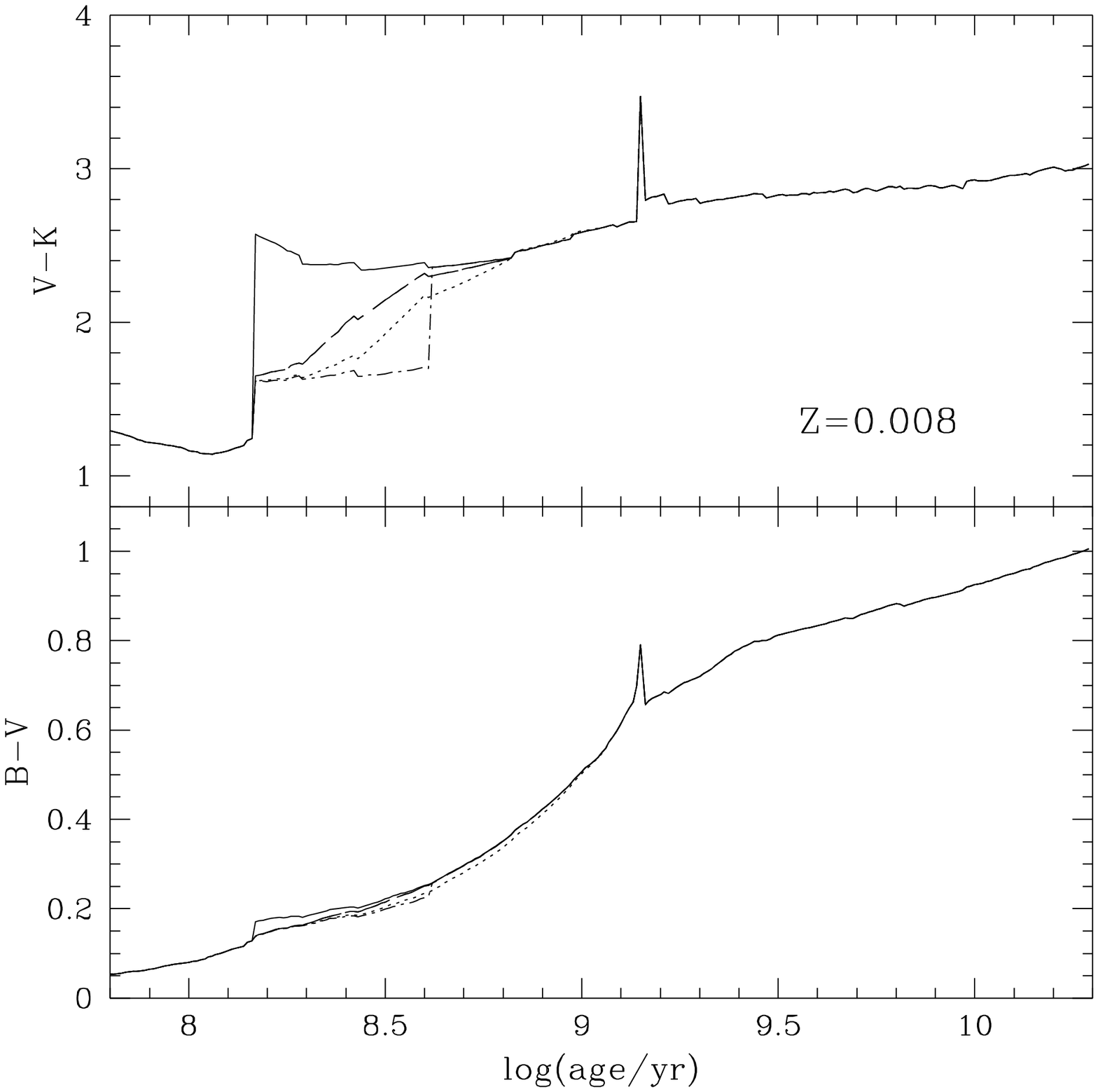,width=8.3cm}
\end{minipage}
        \caption{  
The same as \reffig{fig_colori}, but limited to the case of
Vassiliadis \& Wood's mass-loss rates, and considering the
overluminosity effect on the most massive AGB stars according to 2
different prescriptions: as if the envelope mass were the only
parameter determining the overluminosity [\refeq{eq_overa}; dotted
line], and as if the core mass also plays a role in the overluminosity
effect [\refeq{eq_overb}; long-dashed line]. The continuous line
shows the result for the case in which the overluminosity effect were
ignored. The dotted-dashed line, instead, illustrates the
predictions by Renzini (1992), according to whom the colour jump
associated with the AGB phase transition would be split into two
separate colour jumps. At sufficiently young and old ages, all the
different models follow a common colour evolution.
        }
\label{fig_coloriover}
\end{figure*} 

Figure~\ref{fig_colori} shows, for two metallicity values ($Z=0.001$
and $Z=0.008$), the evolution of the \bv\ and \vk\ colours derived
from a sequence of isochrones calculated with a very fine age spacing,
for the entire age interval in which AGB stars are present. The
several formulations for the mass-loss rates discussed in
\refsec{sec_tpagb} are considered in this plot. These models
correspond to those presented in \reffig{fig_mcore}.
Figure~\ref{fig_coloriover}, instead, shows the colour evolution for
models which assume the Vassiliadis \& Wood's (1993) mass loss rates,
and that consider the effect of overluminosity of the most massive AGB
stars according to the prescriptions discussed in
\refsec{sec_over}. It corresponds to the models shown in
\reffig{fig_mcoreover}.  Most of the considerations below refer to the
results presented in these key figures.

All the SSP models are calculated assuming the simple Salpeter
initial-mass function, $\phi_{M_{\rm i}}\propto M_{\rm i}^{-2.35}$.
As demonstrated in e.g.\ Girardi \& Bica (1993), integrated colours of
SSPs, for ages $t\ga10^8$~yr, depend only weakly on the assumed IMF
slope.

\subsection{The AGB phase transition }
\label{sec_AGBpt}

\subsubsection{The simplest case: results for canonical AGB evolution}

The `AGB phase transition' refers to the almost sudden change in the
evolutionary behaviour of stars, that follows the development of a
electron degenerate CO core prior to the carbon ignition (Renzini \&
Buzzoni 1986). It represents a real bifurcation in the evolutionary
tracks in stars of mass $\Mi\simeq\Mup$: if $\Mi>\Mup$, the star
follows a fast evolutionary history, in which C and O burning proceeds
without a significant increase in \Mcore; if on the contrary
$\Mi<\Mup$, a relatively long evolutionary phase (the AGB) follows
with the possibility of a substantial increase in \Mcore\ before its
end. It results in an increase in the nuclear fuel at a quite precise
value of initial mass.  As this increase occurs only in the form of
red stars, it causes the SSP colours -- specially the reddest ones --
to jump to the red at the age of $t(\Mup)$.

This is the classical view of the AGB phase transition, as idealized
by Renzini \& Buzzoni (1986) and found in a series of SSP models
(e.g.\ Charlot \& Bruzual 1991; Bressan et al.\
1994b). Figure~\ref{fig_colori} illustrates clearly the pronounced
jump in the \vk\ colour that occurs at an age of $t(\Mup)=10^8$~yr,
and how its magnitude depends on the prescription we adopt for the
TP-AGB models. In near-solar metallicity SSPs, the \bv\ colour (as
well as other `visual' colours) is little affected by the onset of the
AGB. This because MS and CHeB phases dominate completely the colour
evolution at these pass-bands. The same does not occur for
low-metallicity SSP models: as the AGB phase develops at relatively
high effective temperatures in this case, also the \bv\ colour is
strongly affected by the development of the AGB, as well as by all the
uncertainties in the AGB evolution.

\subsubsection{The overluminosity affecting the AGB phase transition}

Recently, Renzini (1992) suggested that the appearance of the colour
jump associated with the AGB phase transition could be significantly
delayed, thanks to the overluminosity effect that occurs in the most
massive AGB stars. His argument is essentially that the overluminosity
causes a strong reduction (up to a factor of 10) in the lifetime of
the AGB stars that experience this effect. Thus, the amplitude of the
colour jump at $t(\Mup)$ would be reduced by a similar numerical
factor, and a major jump in the colours would instead occur at a later
age, corresponding to the advent of the first AGB stars that do not
pass through the overluminous phase.  The limit mass for this `normal'
AGB evolution to occur was estimated to be 3~\Msolar, and then the
major colour jump due to the AGB development would occur at ages close
to those of the RGB phase transition (remember that according to
classical models, $\Mhef\simeq2.2$~\Msolar). The situation is also
illustrated in \reffig{fig_coloriover}.

However, Renzini's (1992) analysis does not consider an effect that
became clear in subsequent papers about AGB evolution: that
overluminous stars sooner or later loose enough mass so that they go
down in luminosity and follow a normal CMLR later on (see
\refsec{sec_over}). This behaviour suggests that there is no
clearcut transition from one evolutionary regime to the other (i.e.,
from overluminous to normal) as we vary the initial stellar
masses. On the contrary, the expected behaviour is that the
overluminosity affects the AGB evolution (and hence integrated
colours) less and less as we move from higher to lower initial masses
(and hence from younger to older ages). We can appreciate this smooth
transition in \reffig{fig_mcoreover}, where the overluminosity effect
causes a flattening in the initial--final mass relation, instead of
causing discontinuities in it. The same flattening effect occurs in
the integrated colours, as shown in \reffig{fig_coloriover}.

Therefore, although knowing the limitations of the present models
(\refsec{sec_over}), we propose the interpretation that the
overluminosity with respect to the CMLR causes a {\em temporal
widening} of the colour jump associated to the AGB phase transition,
and not its splitting into two separate events. More specifically, we
suggest that after a colour jump of small amplitude at $10^8$~yr,
there would follow a gradual reddening of the integrated colours up to
ages $\sim3\times10^8$~yr. After that age the overluminosity effect
would not be found in the AGB stars present in the SSPs, and
the colour evolution would follow strictly the curves depicted in
\reffig{fig_colori}. The age of $3\times10^8$~yr corresponds to a
turn-off mass of 3~\Msolar. This is a sort of lower limit to the
values found in the literature (e.g.\ Renzini \& Voli 1981;
Vassiliadis \& Wood 1993; Wagenhuber 1996; Marigo et al.\ 1996a, 1998)
as the limit mass for the presence of envelope burning in TP-AGB
stars.

Comparing the different curves for $(\vk)(t)$, depicted in Figs.\
\ref{fig_colori} and \ref{fig_coloriover}, with the corresponding
curves for $\McoreAGB(t)$ in Figs.\ \ref{fig_mcore} and
\ref{fig_mcoreover}, the reader can notice that the behaviour
predicted by means of the fuel consumption theorem in
\refsec{sec_fct}, is in fact veryfied: the change in the \vk\ colours
due to the presence of the AGB is proportional to the amount of
increase in the core mass during this evolutionary phase. For
instance, for $Z=0.008$ an increase of the core-mass of
$\Delta\Mcore=0.5$~\Msolar\ during the AGB causes an increase of
$\Delta(\vk)=1.6$~mag in the \vk\ colours of the SSPs of corresponding
age. The numerical ratio $\Delta(\vk)/\Delta\Mcore$ is rougly constant
for a given set of metallicities, and equal to 3.2 for $Z=0.008$, and
3.0 for $Z=0.001$. This suggests that
the colour evolution of low-metallicity stellar populations, if
observed, could provide us with useful constraints to the
initial-final mass relationship at other metallicities.  Although it
may seem rather speculative, the star clusters of the SMC would be the
obvious target for a study of this kind.

\subsection{The RGB phase transition }
\label{sec_RGBpt}

Contrarily to the case of the AGB phase transition, the effect of the
development of the RGB on the integrated colours has been debated over
the years.  Renzini \& Buzzoni (1986) argued that a colour jump to the
red should occur, based on the following characteristic features of
the stellar models: (1) the fuel consumption in stars with $\Mi\le\Mhef$
should supposedly increase, due to the development of an extended RGB
phase in which the core mass significantly increases; (2) the advent of
degenerate He cores imply changes on the distribution of the red stars
in the HR diagram, namely the development of the RGB and the increase
of the HB (or CHeB) luminosity by about 0.4 magnitudes. All these
features seem to imply a jump of the integrated SSP colours to the red
at $t(\Mhef)$.

The first complete models of stellar populations to investigate the
behaviour of the colours in the relevant age interval (Chiosi et al.\
1988; Charlot \& Bruzual 1991), brought doubts about the occurrence of
such a colour jump, or at least severely limited its possible
amplitude.  Chiosi et al.\ (1988) and Bressan et al.\ (1994b) argued
that the RGB development is accompanied by a correspondent reduction
in the contribution of the AGB to the integrated light, so that no
significant effect in the colours could result from the increase in
the fuel consumption on the RGB.

\subsubsection{Results from complete SSP models}

Let us now look at the problem with the aid of our models. In
\reffig{fig_colori} the evolution of SSP colors is shown for a
sequence of isochrones calculated with a very small age spacing. In
the vicinity of the age at which the RGB develops,
$t(\Mhef)\simeq10^9$~yr, the following characteristics are noticed:
	\benu 
	\item the colors immediately before and after the onset of the
RGB follow an apparently continuous and regular evolution;
	\label{item_continuous}
	\item there is a small, bump-like excursion to red colors at
an age slightly larger than $t(\Mhef)$.
	\label{item_redphase}
	\eenu

The first characteristic, \ref{item_continuous}, was already described
in the works by Chiosi et al.\ (1988) and Bressan et al.\ (1994b): it
reflects the continuity of the total fuel consumption for stellar
populations which contain AGB stars [see \refsec{sec_fct}, specially
\refeq{eq_totalfuel}].  To be more clear, we recall that what happens
at the onset of the RGB is not a net increase in the total fuel
consumption, but a simple rearrangement of the fuel burned by several
evolutionary phases: a certain amount of fuel that for $\Mi>\Mhef$ was
burned on the CHeB and AGB phases, for $\Mi<\Mhef$ is burned during
the RGB phase.  Most of the rearrangement occurs by the transference
of fuel from CHeB and E-AGB stars to the RGB; the TP-AGB phase
afterwards just levels off the remaining discontinuities in the
function $F\sub{T}(\Mi)$ (see \refsec{sec_core_mass}). And since RGB
and E-AGB stars have virtually the same spectral energy distribution,
the integrated colors do not suffer noticeable changes due to the RGB
phase transition. A result that may be surprising, is that this
behaviour is valid even for the near-IR colors.

The second characteristic -- the bump-like excursion to the red,
\ref{item_redphase} --, as far as we know, was not pointed out
previously.  It can be found only in SSP sequences built with a very
small age spacing.  The first impression one may have is that it could
be a numerical error rather than a real feature of the colour
evolution; we however verified that this is not the case.
Apparently, its occurrence is not predicted from the straight
application of the fuel consumption theorem.
%, nor by the isomass method
%of construction of SSPs.
  
\begin{figure}
\psfig{file=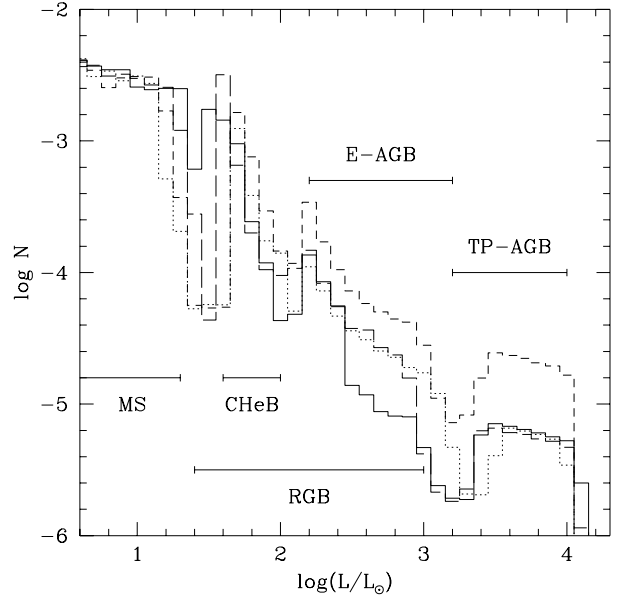,width=8.3cm}
        \caption{  
	The luminosity functions for 4 isochrones with $Z=0.008$. They
have increasing ages of $\log(t/{\rm yr})=9.05$, 9.10, 9.15, and 9.2
(respectively, the continuous, long-dashed, short-dashed and dotted
lines). This age interval goes from the
development of the RGB up to the end of the red phase depicted in
\reffig{fig_colori}. Notice the short RGB present in the 
$\log(t/{\rm yr})=9.05$
isochrone, and the increased number of AGB stars in the 
$\log(t/{\rm yr})=9.15$ one.
        }
\label{fig_lf}
\end{figure} 

In order to clarify the origin of this feature, we present in
\reffig{fig_lf} the theoretical luminosity functions for isochrones
located in the relevant age interval. The great increase in the 
number of AGB stars at the peak of the red phase is remarkable.
This temporary excess number of AGB stars is the main responsible for 
the red phase, and can be understood according to the following.
  
\begin{figure}
\psfig{file=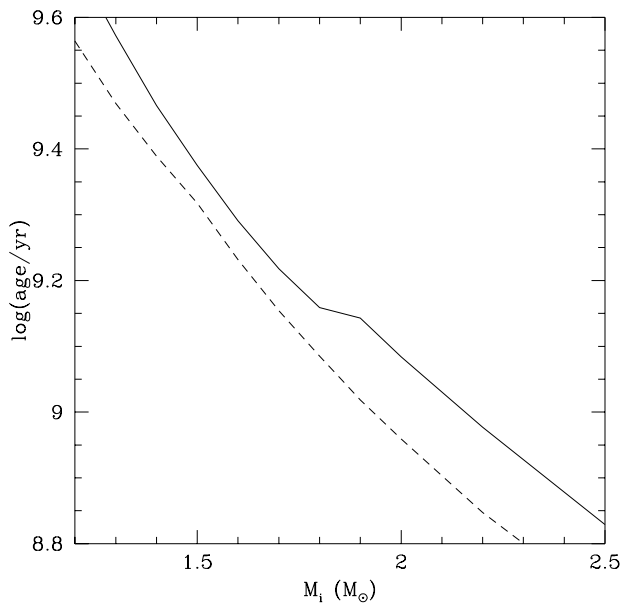,width=8.3cm}
        \caption{  
	The lifetimes of the $Z=0.008$ stellar tracks up to
the end of the main sequence, $t\sub{H}$ (dashed line), and up to the
end of the CHeB, $t\sub{eHe}$ (continuous line). Notice in particular
the flattening of the $t\sub{eHe}(\Mi)$ relation for ages
$\log(t/{\rm yr})\simeq9.1$.
        }
\label{fig_lifetime}
\end{figure} 

The rate of birth of AGB stars in a SSP can be defined similarly to
the evolutionary rate $b(t)$ of \refeq{eq_evrate}:
        \beqa
b^{\rm AGB}(t) & = & \phi_{M_{\rm eHe}} 
         \left| \frac{\diff t\sub{eHe}}{\diff \Mi} 
		\right|_{M_{\rm i}=M_{\rm eHe}}^{-1} \nonumber\\
	& \simeq &
	 \phi_{M_{\rm TO}} 
         \left| \frac{\diff t\sub{eHe}}{\diff \Mi} 
		\right|_{M_{\rm i}=M_{\rm TO}}^{-1} \,.
	\label{eq_evrateAGB}
        \eeqa
where the subscript `eHe' refers to the quantities evaluated at the end
of the CHeB phase. Figure~\ref{fig_lifetime} shows a plot of the
functions $t\sub{H}(\Mi)$ and $t\sub{eHe}(\Mi)$ for our $Z=0.008$
evolutionary tracks, over a limited range of ages and initial masses.
We can notice that, while $t\sub{H}$ is a quite regular function,
$t\sub{eHe}$ flattens at an age of about $\log(t/{\rm yr})=9.15$. It
causes the increase in $b^{\rm AGB}(t)$ at that age, and hence the red
phase in the SSP colours. This particular behaviour of $t\sub{eHe}$
occurs because the He-burning lifetime, $t\sub{He}$, is shortened from
$\sim2\times10^8$ to $10^8$~yr after the RGB develops (see
Table~\ref{tab_z001}).

The age of the red feature does not coincide exactly with the age in
which the RGB first appears. In the $Z=0.008$ models, it occurs at
$9.14\la\log(t/{\rm yr})\la9.16$, while the RGB develops completely
over the age interval $9.03\la\log(t/{\rm yr})\la9.08$. This delay
corresponds to the time lag between the ages in which the stars with
$\Mi=\Mhef$ are either RGB (RGB development) or AGB stars (red phase),
which can be estimated as being similar to the He-burning lifetime, or
$\sim1.5\times10^8$~yr.

We can conclude that a feature in the colour evolution takes place
because the CHeB lifetime $t_{\rm He}(\Mi)$ is strongly reduced in the
vicinity of $\Mi=\Mhef$. This is equivalent to the condition that a
steep gradient in the function $\McoreRGB(\Mi)$ occurs at the onset of
degenerate He-cores (\reffig{fig_mcore}).  Naturally, a steeper
gradient would favour red phases of higher amplitude, but also
correspondly shorter-lived.

We notice, moreover, that it is necessarily a transient effect, which 
may not be confused with the jump to redder colours predicted by
Renzini \& Buzzoni (1986) as a consequence of the RGB development.  In
the framework provided by the fuel consumption theorem, the temporary
red phase we found may be understood as a second-order effect of the
redistribution of the total fuel consumption $F_{\rm T}$ between the
red giant stars present in the SSPs. But no long-lasting increase in
this quantity takes place, therefore no effective colour-jump occurs.

\subsubsection{Further remarks on the RGB phase transition} 

The red phase we found is probably the most important consequence of
the so-called RGB phase transition in the colours of SSPs.  But its
transient nature imposes difficulties for its possible detection in
real stellar populations. In the following we consider two different
cases of astrophysical interest, for which the development of the RGB
was previously claimed to possibly produce important observational
counterparts.  These are (1) the gaps in the color distribution of
Magellanic Cloud star clusters, and (2) features in the
colour/luminosity distribution of galaxies at large redshifts.

We think that this red phase can not cause gaps in the distribution of
integrated colors of star clusters, as it would be the case if it were
a real transition. More probably, it is a source of color dispersion,
mixing some clusters with age $\sim10^9$~yr with their redder (older)
neighbours in two-color planes. The number of clusters that could be
affected by this effect, however, is limited by the relatively short
duration of the red excursion.  As an illustrative case, let us
consider the \ubv\ color distribution of LMC star clusters. In
\reffig{fig_colori} we can notice that for the \bv\ colour and at
$Z=0.008$, the red phase manifests as a small bump in the colour
evolution at ages of 1~Gyr, which amplitude does not exceed
0.05~mag. The work by Girardi et al.\ (1995) indicates that, for a
sample of coeval clusters of similar ages, the intrinsic color spread
due to stochastic effects is of $\sim0.02$~mag in \bv; that due to a
metallicity dispersion of 0.4~dex is of $\sim0.08$~mag, and
$\sim0.1$~mag can be expected from the variations of the foreground
reddening (these values are typical for the rich LMC clusters).  Thus,
the distribution of intrinsic properties of the LMC star clusters
causes a color dispersion which is already higher than that could be
caused by the red phase. The colour dispersion would then completely
mask this small feature in the colour evolution. For this reason, we
think that gaps in the distribution of \bv\ colors, as e.g.\ the one
noticed by Bica et al.\ (1991), probably can not be formed as a
consequence of the red phase. The same holds for the near-infrared
colours, since for them the colour dispersion due to stochastic
effects gets very high (e.g.\ Santos Jr.\ \& Frogel 1997).

In a galactic population, two of the possible factors of color
dispersion above considered disappear: (1) the stellar statistics is
much higher, and (2) the reddening, either internal or foreground, is
expected to have a smooth distribution, affecting equally all the
components of the population. Even in this case, the red phase would
reflect in the evolution of the galactic colours only if
        \begin{enumerate} 
        \item the components of the galactic population were coeval,
to within $\sim10^8$~yr; (Probably not even the oldest elliptical
galaxies are expected to have formed on so short timescales.)
        \item the red phase were to appear at equal ages in SSPs of
different metallicities, within the typical metallicity interval that
composes a galaxy.
        \end{enumerate} 
As none of these condictions is likely to occur, we conclude that the
red phase does not provide a useful `evolutionary clock', in the sense
idealised by Renzini \& Buzzoni (1986).

\section{The near-infrared evolution observed in LMC clusters}
\label{sec_lmc}

The rich star clusters of the Magellanic Clouds provide the most
significant data for testing SSP models. They span a wide range of
ages, being at the same time rich enough so that stars in some of the
fast evolutionary stages are sampled. Moreover, for a reasonable
number of these clusters, metallicities and ages have been determined
with a good precision, by means of both spectroscopy of the red giant
stars and CCD photometry down to the main sequence. For the large
number of clusters for which only integrated photometry is available,
ages can be estimated by means of the observed relationships between
integrated photometric parameters and age, calibrated with the data
for the best studied clusters.  The most popular among these methods
are the SWB classification scheme of Searle, Wilkinson \& Bagnuolo
(1980), based on $uvgr$ photometry, and the $S$ parameter of Elson \&
Fall (1985), based on $UBV$ photometry.

Persson et al.\ (1983) observed 84 rich star clusters of both
Magellanic Clouds in integrated $JHK$ photometry. The combination of
this data with the $UBV$ photometry of other authors (see van den
Bergh 1981; Bica et al.\ 1996) provides the basic dataset for the study
of the evolution of the \vk\ colour. Persson et al.\ were the first to
call attention to the apparent jump of the \vk\ colour from values
$\sim1.5$ to $\sim3$, occurring for star clusters of SWB type IV. This
remains, basically, the main aspect of the data to be described by SSP
models.

Other characteristic of the observational data was the large
dispersion in the IR colours for clusters of the same SWB type (and
hence similar age).  Although part of this dispersion can be
attributed to observational errors, it is clear that most of it
derives from the natural dispersion in the number of evolved stars
present in the clusters. In the theoretical models, this effect can be
easily simulated by means of stochastic fluctuations in the initial
mass function of model clusters (cf.\ Barbaro \& Bertelli 1977; Chiosi
et al.\ 1988; Girardi et al.\ 1995; Santos Jr.\ \& Frogel 1997). These
effects on the dispersion of near-IR colours of Magellanic Clouds
clusters have been recently studied by Santos Jr.\ \& Frogel
(1997). In this paper, instead, we focus on the study of the mean
evolution of the colours as a function of age.

\subsection{The data about cluster ages}

In order to study the evolution of the near-infrared colours of MC
clusters, we must first properly rank the clusters observed by Persson
et al.\ (1983) into an age sequence. The latter authors based their
own age ranking on a subdivision of the SWB classification (Searle et
al.\ 1980).  More precisely, they drew over the $Q_{ugr}\times
Q_{vgr}$ diagram of Searle et al.\ (1980; see Fig.~1 in Persson et
al.\ 1983) a mean ridge line through the distribution of star
clusters, scaling it by means of an age parameter $t_{\rm PACFM}$ from
0 to 300. Then, a value
of $t_{\rm PACFM}$ was attributed to each cluster by projecting it
over the mean line. The assumption that $t_{\rm PACFM}$ represents
really an age sequence, derives from the demonstration by Searle et
al.\ (1980), that their mean SWB groups (also defined in the
$Q_{ugr}\times Q_{vgr}$ diagram) represent a sequence of increasing
age and decreasing metallicity.

Another approach is that followed by Elson \& Fall (1985). They have
drawn a similar ridge line along the distribution of star clusters in
the integrated \ubbv\ plane, attributing to each cluster an age
parameter $S$. The relationship between $S$ and the logarithm of the
age was found to be linear, and was calibrated by using a large amount
of photometric data (see also Elson \& Fall 1988). Girardi et al.\
(1995) have revised the method, basically confirming its validity. Two
modifications were however suggested in the way of attributing the $S$
parameter, aimed at more realistically represent the age sequence of
the oldest clusters, as well as those falling far from the mean $S$
ridge line.

In our opinion, a revision of the age classification adopted by
Persson et al.\ (1983) would also be worth, for two reasons.  First of
all, the age calibration of SWB groups is based on old data, of very
low quality if compared to those available nowadays. Second, the SWB
classification, as well the behaviour of integrated $uvgr$ colours,
has never been analised with the aid of SSP models of different ages,
metallicities, and number of stars. In other words, it has not been
demonstrated that the distribution of clusters in the $Q_{ugr}\times
Q_{vgr}$ diagram provides a good-quality age indicator. At least one
of the assumptions of Persson et al.\ (1983), that the scattering
along the ridge line is due only to observational errors, can be
questioned by means of very simple simulations of star clusters. In
fact, the stochastic dispersion of LMC cluster properties is not
negligible in the visual colours, and more importantly, it occurs
along preferential directions in colour-colour diagrams (see the
analysis of Girardi et al.\ 1995 for $UBV$). It represents a potential
origin of both uncertainties and systematic errors in the methods for
age attribution based on colour-colour diagrams.

\reftab{tab_ir_lmc} presents the value of the $S$ and $t_{\rm PACFM}$
age indicators, for those LMC clusters for which the $JHK$ photometry
is available.  In the first column we have the parameter $S$ as
defined by Girardi et al.\ (1995) and according to the $UBV$ data from
Bica et al.\ (1996). It follows in the second column, the $S$ as
defined originally by Elson \& Fall (1985), $S\sub{EF}$. We can notice
that both values are very similar, since the listed clusters are in
general bright ones, falling close to the ridge $S$ line drawn in both
works. We recall that the $S$ parameter is shown to correlate with age
according to $\log(t/{\rm yr})=6.227+0.0733\,S$, giving a r.m.s.\
precision of 0.14~dex in the age determination of rich clusters
(Girardi et al.\ 1995). However, according to the latter authors, it
can not be used to date old clusters of low metallicity (those with
SWB type VII in the Magellanic Clouds). Therefore, these are assumed
to have 15~Gyr, or equivalently $S=54.5$.  The $t_{\rm PACFM}$ from
Persson et al.\ (1983) is listed in the third column of
\reftab{tab_ir_lmc}.  It is essentially a non-calibrated age
parameter.

\begin{figure}
\psfig{file=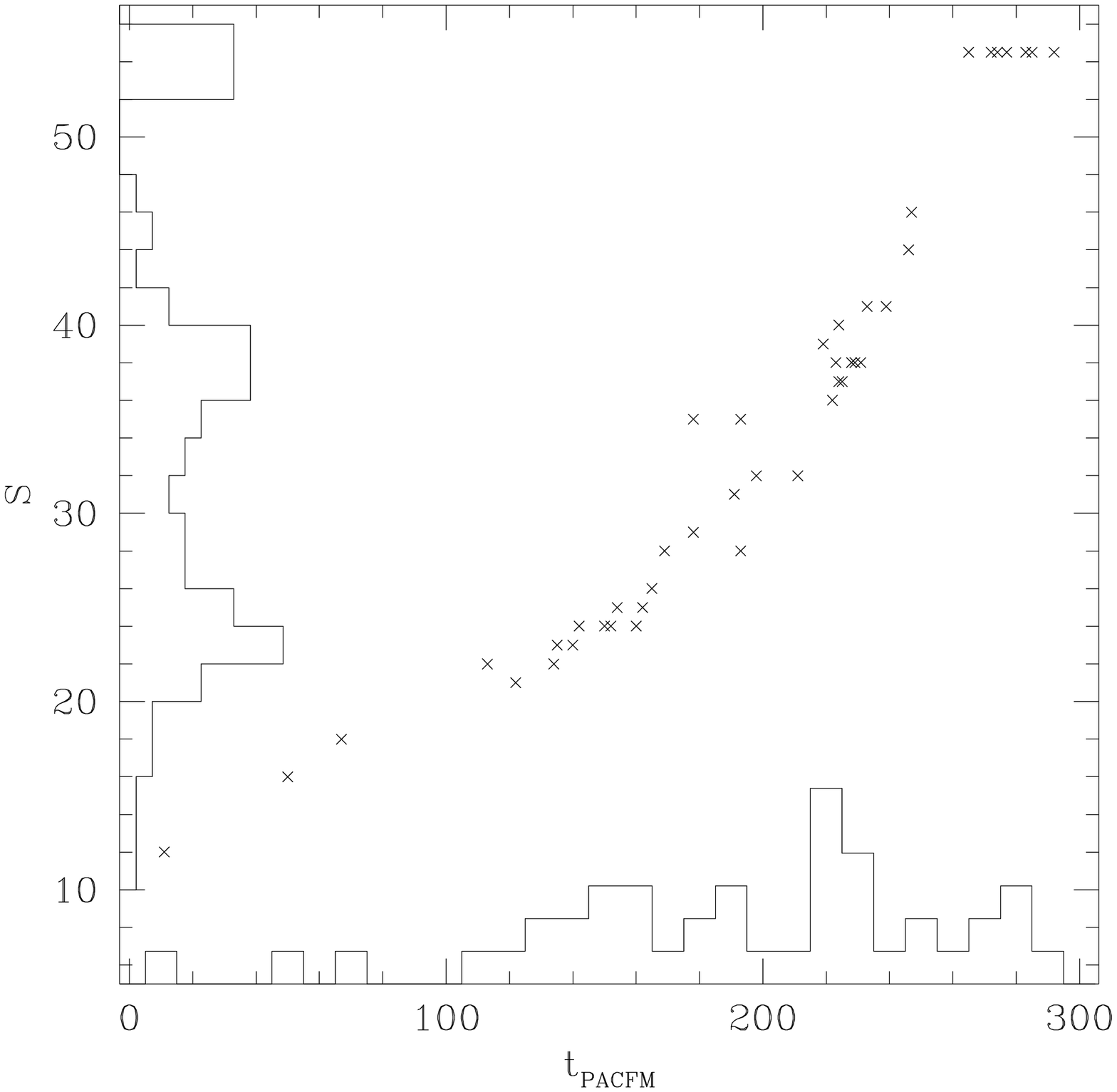,width=8.3cm}
        \caption{ 
Relationship between the parameters $S$ obtained by Girardi et al.\
(1995), and $t_{\rm PACFM}$ from Persson et al.\ (1983), for the LMC
clusters of \reftab{tab_ir_lmc} which present both age indicators
(crosses). The histograms along the two axes give the number
distribution of both parameter. See the text for more details.
	}
\label{fig_ages}
\end{figure} 

\begin{table*}
\caption{Age indicators for LMC clusters with near-IR photometry.}
\label{tab_ir_lmc}
\begin{tabular}{llllllllllll}
\noalign{\smallskip}\hline\noalign{\smallskip}
Id.\ (NGC) & $S$ & $S\sub{EF}$ & $t\sub{PACFM}$ & 
Id.\ (NGC) & $S$ & $S\sub{EF}$ & $t\sub{PACFM}$ & 
Id.\ (NGC) & $S$ & $S\sub{EF}$ & $t\sub{PACFM}$ \\
\noalign{\smallskip}\hline\noalign{\smallskip}
1466 &  -- &  49 & 274 & 1872 &  29 &  30 & 178 & 2108 &  36 &  36 &  -- \\
1644 &  37 &  37 & 224 & 1898 &  -- &  50 &  -- & 2121 &  46 &  44 & 247 \\ 
1711 &  21 &  20 & 122 & 1916 &  -- &  46 &  -- & 2134 &  28 &  28 & 193 \\ 
1751 &  40 &  42 & 224 & 1943 &  24 &  25 & 150 & 2136 &  25 &  26 & 154 \\
1755 &  24 &  24 & 142 & 1944 &  29 &  29 &  -- & 2154 &  38 &  39 & 229 \\
1767 &  14 &  16 &  -- & 1951 &  25 &  24 &  -- & 2155 &  44 &  45 & 246 \\
1774 &  22 &  23 & 113 & 1953 &  31 &  29 &  -- & 2156 &  28 &  26 &  -- \\
1783 &  37 &  37 & 225 & 1978 &  41 &  45 & 239 & 2157 &  25 &  25 &  -- \\
1786 &  -- &  48 &  -- & 1984 &  10 &  11 &  -- & 2159 &  24 &  25 &  -- \\
1805 &  17 &  17 &  -- & 1986 &  23 &  24 & 135 & 2162 &  36 &  39 & 222 \\
1806 &  38 &  40 & 231 & 1987 &  35 &  35 & 193 & 2164 &  24 &  23 & 160 \\
1818 &  18 &  18 &  67 & 1994 &  13 &  15 &  -- & 2172 &  25 &  25 &  -- \\
1831 &  32 &  31 & 211 & 2002 &  13 &  17 &  -- & 2173 &  41 &  42 & 233 \\
1835 &  -- &  47 & 272 & 2004 &  12 &  15 &  11 & 2209 &  35 &  35 & 178 \\
1841 &  -- &  42 & 292 & 2011 &  13 &  13 &  -- & 2210 &  -- &  48 & 277 \\
1846 &  39 &  40 & 219 & 2019 &  -- &  46 & 265 & 2213 &  38 &  39 & 228 \\
1850 &  23 &  21 &  -- & 2041 &  24 &  25 & 152 & 2214 &  23 &  22 & 140 \\
1854 &  22 &  24 & 134 & 2058 &  25 &  26 & 162 & 2231 &  38 &  37 & 223 \\
1855 &  22 &  22 &  -- & 2065 &  26 &  26 & 165 & 2257 &  -- &  51 & 285 \\
1856 &  31 &  30 & 191 & 2070 &  18 &  14 &  -- & H~11 &  -- &  51 & 283 \\
1866 &  28 &  27 & 169 & 2100 &  16 &  17 &  50 &      &     &     &     \\
1868 &  33 &  33 &  -- & 2107 &  32 &  32 & 198 &      &     &     &     \\
\noalign{\smallskip}\hline\noalign{\smallskip}
\end{tabular}

\smallskip Clusters with no $S$ parameter are of SWB type VII,
and then supposedly old ($t\ga12$~Gyr).\\
\end{table*}

Figure~\ref{fig_ages} presents the relation between the parameters $S$
and $t_{\rm PACFM}$, as well as histograms of their values, for the
clusters of \reftab{tab_ir_lmc}. It is clear that the relation is not
linear, and that the clusters tend to concentrate at certain values of
age parameters. For instance, $S$ values around 23 and 38, and $t_{\rm
PACFM}$ values around 150 are 220, are more frequent between the
clusters. In Girardi et al.\ (1995), these peaks in the distribution
of $S$ are interpreted as being the signature of different episodes of
enhanced cluster formation in the LMC. The histogram of $t_{\rm
PACFM}$, however, presents a peak around 190, and a minimum abround
210, which have no counterparts in the distribution of $S$.
This difference probably reflects the low number of objects in 
the sample. We notice also that $t_{\rm PACFM}=200$ is
exactly the transition value between clusters with $\vk\simeq1.5$ and
3 (see Fig.\ 2 in Persson et al.\ 1983, and \reffig{vk_lmc} below). It
corresponds, also, to a period of apparently fast evolution of \bv\
colours, according to the latter authors.

\subsection{The transition of \vk\ colours}

\begin{figure}
\psfig{file=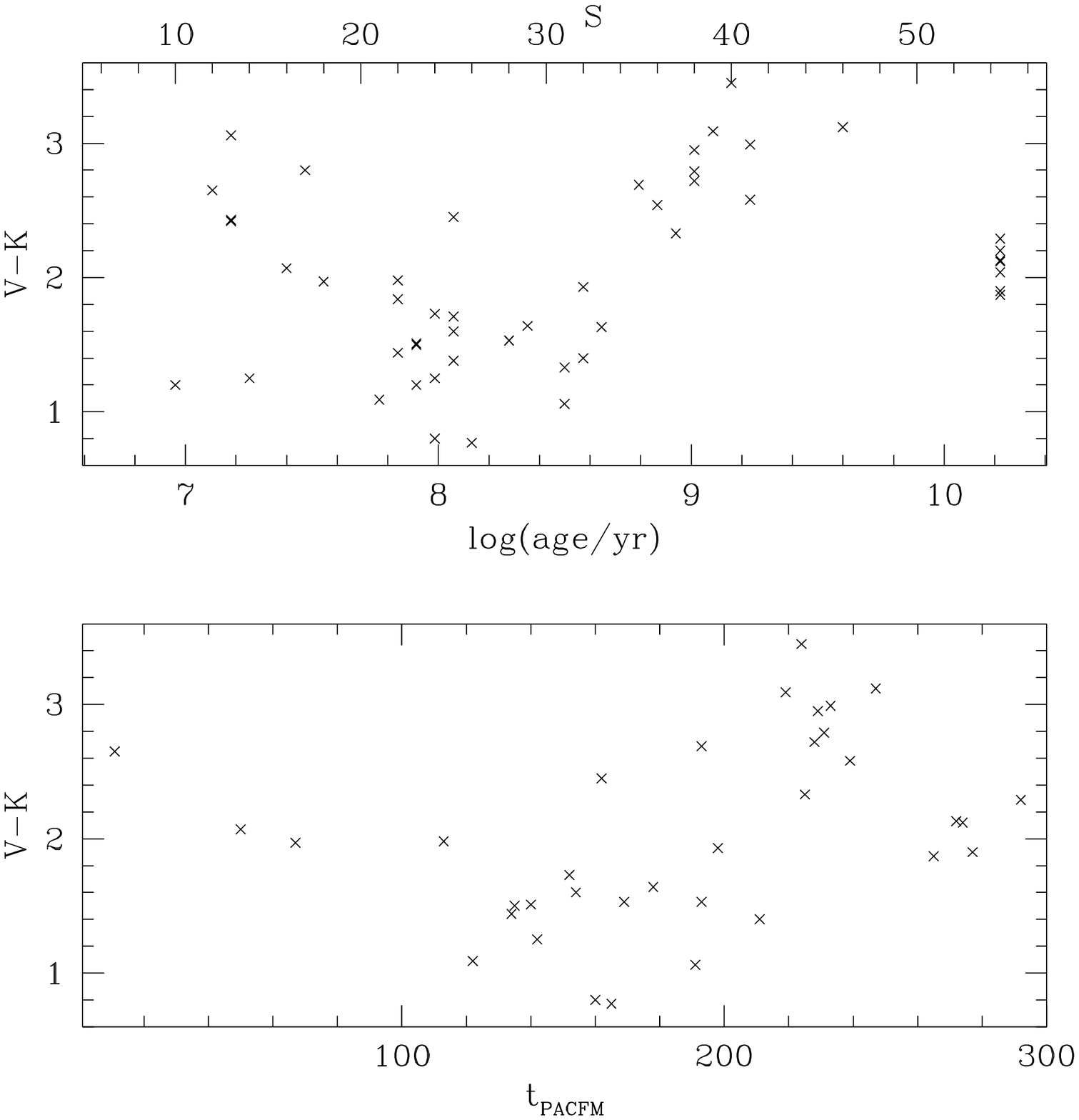,width=8.3cm}
        \caption{ 
The time evolution of \vk\ for LMC clusters, according to two
different age rankings: the $S$ parameter as defined and calibrated by
Girardi et al.\ (1995; upper panel), and the $t_{\rm PACFM}$ parameter
from Persson et al.\ (1983; lower panel).
	}
\label{vk_lmc}
\end{figure} 

Figure~\ref{vk_lmc} shows the age distribution of \vk\ colours of LMC
star clusters, when we use the ages derived from the $S$ parameter as
in Girardi et al.\ (1995), or the age indicator $t_{\rm PACFM}$.  We
remark that there are many more clusters with ages determined by means
of the former method.

In the upper panel we notice several interesting
characteristics. First, for ages slightly larger than $10^7$~yr, we
have clusters with \vk\ colours spanning the range
$1\la(\vk)\la3$. For those with \vk\ close to 3, the red colours can
be explained by the presence of red supergiant (RSG) stars in clusters
of similar age (see e.g.\ Arimoto \& Bica 1989; Girardi \& Bica
1993). Classical examples of these clusters are NGC~2004 and 2100 (see
Bica, Alloin \& Santos Jr.\ 1989). The youngest cluster in the sample
(NGC~1834, with $S=10$ and $\vk\simeq1.2$) is probably too young to
contain red supergiant stars.

At ages around $10^8$~yr, cluster colours concentrate around
$\vk\sim1.5$, while at $10^9$~yr they have $\vk\sim3$. The velocity of
this colour transition, due to the high dispersion of the colours and
to the small number of clusters, can not be precisely determined.
However, comparison of the two panels in \reffig{vk_lmc} reveals that
the $S$ age-ranking allows either for a fast colour transition at
$\simeq34$, or for a {\em gradual} colour transition occurring in the
interval $30<S<40$, while the $t_{\rm PACFM}$ one would suggest
(according to the interpretation originally given by Persson et al.\
1983, see their Fig.\ 2), a fast colour transition occurring at
$t_{\rm PACFM}=200$.

Therefore, the fast transition in \vk\ is not so clearly defined in
the plot which uses the $S$ parameter, which is arguably a better age
indicator than $t_{\rm PACFM}$. We recall that the adoption of $t_{\rm
PACFM}$ as an age indicator allows also for a relatively fast
transition of the \bv\ colours (see Fig.\ 3 in Persson et al.\ 1983)
at almost the same age of the presumed \vk\ transition (see also Corsi
et al.\ 1994). There is no sign of such a feature, however, in plots
that relate the $S$ parameter (and in turn of the \bv\ colour) with
cluster ages determined directly from the main-sequence turn-off
magnitude (see Elson et al.\ 1988; and Fig.\ 11 in Girardi et al.\
1995).  Therefore, we point out that the fast transition of \vk\ at
$t_{\rm PACFM}\sim 200$, could simply reflect a deficiency of this
parameter in representing the age sequence of star clusters, instead
of a real feature of the \vk\ evolution.

Finally, we notice that in the upper panel of \reffig{vk_lmc}, the old
clusters (with assumed ages of 15~Gyr) have $\vk\sim2$, being bluer
than those with $\sim10^9$~yr. This probably reflects their lower
metallicity.  According to Olzewski et al.\ (1991), the old clusters
in the LMC have $-1.5>\feh>-2.2$, in contrast with the young and
intermediate-age ones, which have $0.0>\feh>-0.6$.

\begin{figure}
\psfig{file=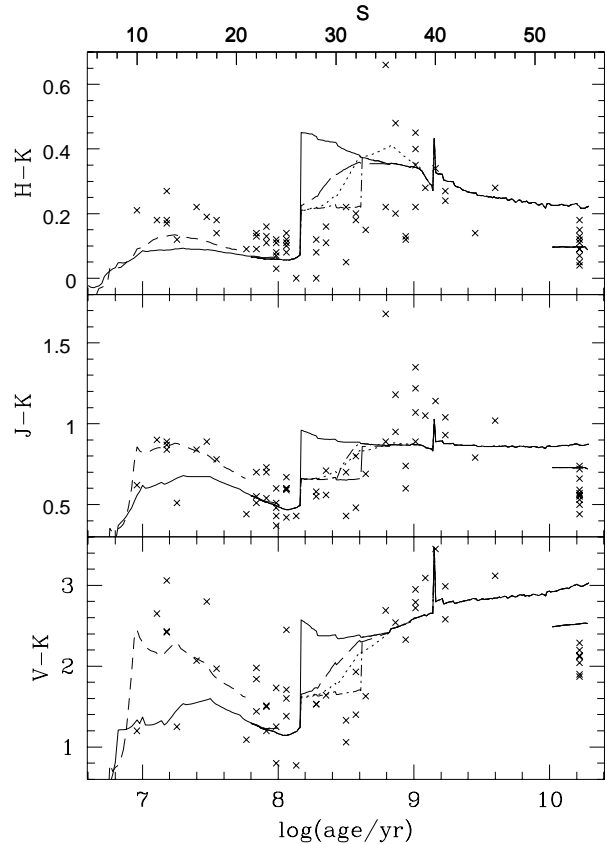,width=8.3cm}
        \caption{ 
Data on the near-IR colours of LMC star clusters, compared to models
of colour evolution with $Z=0.008$, which assume the Vassiliadis \&
Wood (1993) prescription for the mass-loss rates.  The meaning of the
lines is as in \reffig{fig_coloriover}. Models with $Z=0.001$ are
shown only for $t>10$~Gyr. For $\log(t/{\rm yr})<7.8$ we present
also the case $Z=0.02$ from Bertelli et al.\ (1994; short-dashed line).
	}
\label{fig_ir_lmc}
\end{figure}

Figure~\ref{fig_ir_lmc} presents the complete data for the \vk, \jk,
and \hk\ colours, for ages derived from the $S$ parameter.
Superimposed, are the SSP models for $Z=0.008$, and calculated with
the Vassiliadis \& Wood's (1993) mass-loss rates (cf.\
\reffig{fig_coloriover}). For $\log(t/{\rm yr})<7.8$, we draw also the 
models of Bertelli et al.\ (1994) for $Z=0.02$. The latter
models are important for a discussion of the colours of clusters which
contain RSG stars. We notice that models with $Z=0.02$ reproduce well
the location of the very young [$\log(t/{\rm yr})<7.8)$] and red
clusters. Those with $Z=0.008$, instead, are too blue to describe most
of the data in this age interval. This is probably related to the fact
that theoretical models with $Z=0.008$, in general, predict too blue
RSGs with respect to those observed in LMC clusters. Other effects, as
the sensitivity of the colours to the IMF slope (e.g.\ Girardi \& Bica
1993), may also affect the colours in this age range around $10^7$~yr.

With respect to the \jk\ and \hk\ colours, we recall that they refer
only the form of the near-infrared spectrum.  They are very sensitive
to the presence of carbon stars in a stellar cluster (Persson et al.\
1983; Frogel, Mould \& Blanco 1990).  The red \jk\ and \hk\ colours
that occur in the clusters with age $8.5\la\log(t/{\rm yr})\la9.2$, in
fact, can be attributed to the presence of carbon stars in this
limited age range (see Frogel et al.\ 1990; Marigo et al.\ 1996b). As
our models of SSPs do not differentiate M- and C-type AGB stars, they
fail in giving a good description of these colours. This can be seen
in the upper and middle panels of \reffig{fig_ir_lmc}.

The lower panel, with the evolution of \vk, is far more illustrative.
It shows that the $Z=0.008$ models for sufficiently young and old
ages, reproduce the observed clumping of colours at values
$\vk\sim1.5$ at $10^8$~yr, and $\vk\sim3$ at $10^9$~yr. The oldest
models with $Z=0.001$ approximate to the colours $\vk\sim2$ observed
for the oldest clusters. Anyway, this value of the metallicity is yet
too high for allowing a significant comparison with the data for the
old clusters (values of $Z$ as low as 0.0005 would be required).

The most interesting comparison arises in the age range from $10^8$ to
$10^9$~yr, in which the \vk\ colour transition takes place. We can see
that models which assume no overluminosity for the AGB stars (i.e.,
`classical' models), present too red \vk\ colours in the first part of
this age interval. Models which include this effect, instead,
approximate to the mean locus of the observed points. Also the model
which represents the prediction by Renzini (1992) -- which assume a
more dramatic prescription for the evolution of envelope burning stars,
see \refsec{sec_AGBpt} -- would describe better the observed points than
classical models. Unless the assumed prescriptions for the TP-AGB
evolution are very far from the reality, {\em this can be seen as an
indication for the occurrence of overluminosity (and hence envelope
burning) in the AGB stars of the LMC with initial masses $\ga3$~\Msolar}.

\subsection{The role of the RGB}

Another point that we can observe in \reffig{fig_ir_lmc} is that the
RGB appears in the SSP models at quite large ages [at $\log(t/{\rm
yr})=9.08$, or $S=39$], when the transition from $\vk\sim1.5$ to 3.0
is practically completed. Therefore, the development of the RGB has no
role in this transition. The development of the AGB, instead, is
determinant. This can be observed in the right panel of
\reffig{fig_colori}: were the TP-AGB absent in our models [as is
practically the case when we use the Bl\"ocker's (1995) mass-loss
rates], the transition of the \vk\ colour would also occur, but much
more slowly. Values of $\vk\simeq3$ would be attained only at ages of
$\sim10^{10}$~yr in this case, and not at $\sim10^9$~yr as observed.

This is not in contradiction with the results obtained from Corsi et
al.\ (1994) and Ferraro et al.\ (1995), which conclude that the
development of the AGB is the main responsible for the \vk\
transition. However, they suggest that the RGB appears in the
Magellanic Cloud clusters with $S>35$, i.e.\ practically in the age
range in which the \vk\ colour more rapidly change (see
\reffig{vk_lmc}), contrarily to our results.  It is worth comparing
our models with their data, in order to identify possible
discrepancies.

\begin{figure}
\psfig{file=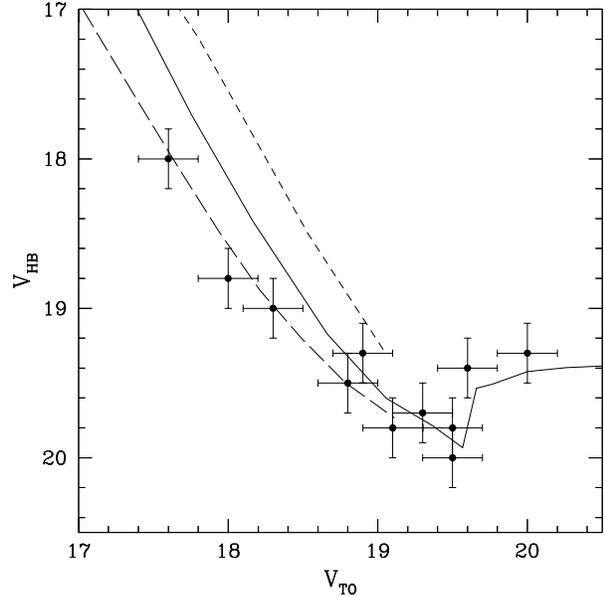,width=8.3cm}
        \caption{ 
The magnitude of the clump of CHeB stars, $V\sub{HB}$, against the
terminal-age main sequence, $V\sub{TO}$, as observed in LMC star
clusters by Corsi et al.\ (1994; black dots) and as predicted by our
$Z=0.008$ models (continuous line). The data are presented with the typical
error bars of 0.2~mag in $V$. An apparent 
distance modulus of $V-M_{V}=18.7$~mag was
assumed for the LMC. Long-dashed lines represent the models for $Z=0.02$ 
from Bertelli et al.\ (1994), for $\log(t/{\rm yr})<9$, also computed with
moderate convective overshoot. The 
short-dashed line presents the corresponding relation obtained from
Alongi et al.\ (1993) models with $Z=0.008$, but assuming the
semiconvective scheme of mixing.}
\label{fig_mv_to_hb}
\end{figure} 

These authors deal basically with two complementary data sets.  Corsi
et al.\ (1994) obtained $BV$ CCD photometry for 11 LMC clusters with
intermediate values of $S$. They determine quite precisely the
location of the terminal main sequence ($V\sub{TO}$) and the red clump
of He-burning stars (or $V\sub{HB}$) for each cluster. A plot of
$V\sub{TO}$ against $V\sub{HB}$ (their Fig.~36, and
\reffig{fig_mv_to_hb} below) reveals clearly the presence of a minimum
in the function $V\sub{HB}(V\sub{TO})$, for clusters around NGC~2209,
which have $S=35$. This minimum reflects the similar minimum in
$\McoreRGB(\Mi)$ present in the stellar evolutionary models
(\reffig{fig_mcore}), and should occur at ages immediately before the
onset of the RGB. The two clusters observed in the plateau of the
$V\sub{HB}(V\sub{TO})$ curve, corresponding to clusters which have
already well-developed RGBs, have $S\sub{EF}=39$ and 43 (Elson \& Fall
1985), or $S=36$ and 41 (Girardi et al.\ 1995).

In \reffig{fig_mv_to_hb} we present a similar plot of $V\sub{TO}$
versus $V\sub{HB}$, comparing the Corsi et al.\ (1994) data with the
results from our $Z=0.008$ models (continuous line). 
The latter are displayed in the diagram after shifted by a 
quantity corresponding to the mean apparent
distance modulus of the LMC clusters, $V-M_{V}=18.7$. 
These models seem to describe well the general distribution of points,
especially in the minimum of the $V\sub{HB}(V\sub{TO})$ curve. 
However, for the youngest clusters, 
the observed $V\sub{HB}$ is systematically shifted to higher
magnitudes with respect to the models. It is not easy to explain this
shift. The adoption of a different distance modulus for
the LMC, for instance, would change the situation only for those
two clusters with $V\sub{TO}>19.5$, the oldest in the sample. 

Several factors could affect
the relative $V$ magnitude of TO and CHeB stars in the young clusters. 
Two of them are also illustrated
in \reffig{fig_mv_to_hb}: (1) Models with higher metal and/or helium content tend
to reproduce better the points with $V\sub{TO}<19$, as shown by the 
Bertelli et al.\ (1994) models with $[Z=0.02, Y=0.28]$ (long-dashed line in 
\reffig{fig_mv_to_hb}). (2) Models with more efficient convective overshoot 
also tend to present higher $V\sub{HB}$ magnitudes for a given $V\sub{TO}$,
(which would also lead to a better agreement with the data).
This can be seen if we compare in \reffig{fig_mv_to_hb} the models with
$[Z=0.008, Y=0.25]$ of Alongi et al.\ (1993) computed with the classical 
semiconvection scheme (short-dashed line), with our models obtained with the 
same chemical composition, but with the adoption of moderate convective overshoot
(continuous line).
Moreover, other effects as systematic errors in the bolometric
corrections used to convert the theoretical quantities to the
magnitude \mv; or
the blending or presence of double stars at the tip of the main
sequence (see e.g.\ Maeder 1994), could also affect the quantities 
presented in \reffig{fig_mv_to_hb}. 
Given the complexity of factors in play, we limit ourselves to 
the comment that the figure points to a good agreement between theory
and observations in which concerns the 
behaviour of $V\sub{HB}(V\sub{TO})$ curve in the vicinity of \Mhef. 

Ferraro et al.\ (1995) have observed in $JHK$ photometry a subsample
of the clusters studied by Corsi et al.\ (1994). RGB stars were
observed for clusters with $S>35$, but unambiguosly only in those with
$S>37$.  They recognize that the present data, due to the problems of
field contamination and small stellar statistics, should be improved
in order to determine more precisely the age in which the RGB stars
appear in the CMDs.  Particularly unfortunate was the fact that the
CMDs of the two clusters they observed in the range $35<S<37$,
NGC~1987 and NGC~2108, can be severely contaminated by the LMC bar
population.

Therefore, the $S$ limit for the appearance of extended RGBs seems to
be located somewhere between $S=35$ and 39, being $S=37$ a more
probable value.  In our models, the RGB develops completely only at
ages corresponding to $S=39$.  Is the difference between these two
results significant\,?  An answer to the above question would require
a detailed revision of the adopted $S$ vs.\ $\log({\rm age})$
calibration, which is beyond the scope of this paper.  Suffice it to
say that systematic shifts in \bv\ colours, with respect to those
observed in clusters, are possibly present in the models. These shifts
probably amount to only $\Delta(\bv)\simeq0.1$ (corresponding to
$\Delta S\simeq2$ in the age range here in consideration), and can be
easily explained by the presence of inadequancies in the theoretical
colour-$\Teff$ relations we used. Moreover, we recall that the
relationship between the $S$ parameter and age can be affected by a
particularly insidious problem in the interval $30\la S\la36$: in this
age range the intrinsic dispersion of cluster colours runs almost
parallel to the $S$ sequence in the \ubbv\ diagram (cf.\ Girardi et
al.\ 1995). This determines an higher uncertainty in the ages of
clusters found around this region of $S$ parameter. It also suggests
that, in future analysis of the subject, ages determined directly from
main sequence photometry (or simply the $V\sub{TO}$ quantity) should
be preferred in order to provide an age ranking to the clusters.

We conclude that there is no fundamental contradiction between the
conclusions obtained in this paper, and those from Corsi et al.\ (1994)
and Ferraro et al.\ (1995). Both theory and observation can probably be
improved in order to obtain a better agreement about the age of the
onset of the RGB, but anyway the present disagreement, if any, is of 
modest amount.

We would like to stress also the necessity of more data concerning the
Magellanic Cloud clusters, in order to provide conclusive checks to
the current SSP models, in the age ranges they can be more
uncertain. It includes obviously the age range from $10^8$ to
$10^9$~yr, in which the main \vk\ transition occurs.  Projects as
those pursued by Corsi et al.\ (1994) and Ferraro et al.\ (1995)
should certainly continue, in order to provide data for more clusters,
so reducing the uncertainties related to the statistics of the gians
stars. Moreover, the question of field contamination in the observed
CMDs has revealed to be crucial in the observational studies carried
out so far.

\section{Final comments}
\label{sec_conclu} 

This work shows that the colour (and hence spectral) evolution of the
SSPs that contain AGB stars, is not as well established as generally
supposed.  Uncertainties in the theory of AGB evolution, regarding the
dependency of mass loss rates on stellar parameters, and the effect of
overluminosity with respect to the core mass--luminosity relation, are
able to significantly modify the results obtained so far for SSPs with
ages from about $10^8$ to $\sim10^9$ yr. Also, the present models
allow for a revision of the colour changes that follow the AGB and RGB
developments. On the one hand, we argue that the colour transition due
to the AGB development should be more gradual that classically
predicted, thanks to the effect of AGB overluminosity.  On the other
hand, we confirm earlier suggestions that the RGB development does not
cause a transition to redder colours, but indicate that a transient
red phase should occur due to the reduction of the core-He burning
lifetimes that accompanies the RGB development. The latter effect,
however, probably does not produce significant observational
counterparts.

The effect of overluminosity in the most massive AGB stars should be
properly included in SSP calculations, in order to define precisely
how the colour evolution occurs for different metallicities. This
overluminosity is expected to affect the near-infrared colours of
near-solar SSPs, and most colours of low-metallicity SSPs, for ages
between $\sim10^8$ and $\sim3\times10^8$ yr. In fact, we show that the
overluminosity provides a reasonable description for the observed
colour evolution of \vk\ colours, from values $\sim1.5$ to 3, in LMC
clusters of these ages. We notice also that envelope burning (and
overluminosity) in TP-AGB stars is thought to be particularly
efficient at low-metallicities. Therefore, this effect is potentially
important in describing the colour evolution of young low-metallicity
stellar populations, even at visual pass-bands.  All these aspects may
be significant in view of realistically predicting the colour
evolution of low-metallicity and young galaxies, observed at high
redshifts.

\section*{Acknowledgments}

L.\ Girardi acknowledges the constructive interaction with E.\
Bica, A.\ Bressan, C.\ Chiosi, P.\ Marigo, and E.\ Nasi, during the
last few years, which made possible the development of this work.
Thanks are due to A.\ Schmidt and M.V.\ Copetti for their
efforts in providing him with suitable working conditions during a
preliminary stage of this research; to the Department of Astronomy of
the Padova University, and the Instituto de F\'\i sica da Universidade
Federal do Rio Grande do Sul for the repeated hospitality; and to the
Alexander von Humboldt Foundation for financial support. A.\ Weiss
and M.\ Groenewegen are acknowledged for the critical reading of the
manuscript.

\label{lastpage}

\end{document}